\documentclass[journal=ancac3,manuscript=article]{achemso}

\usepackage[version=3]{mhchem} % Formula subscripts using \ce{}
\usepackage{color}
\author{B. Shadrack Jabes}
\affiliation{Institute for Mathematics, Freie
Universitat Berlin, D-14195 Berlin, Germany}
\email{shadrack.b@fu-berlin.de (B. S. J)}
\author{L.Delle Site}
\email{luigi.dellesite@fu-berlin.de (L.D.S)}
\affiliation{Institute for Mathematics, Freie
Universitat Berlin, D-14195 Berlin, Germany}

\title[Nanoscale supra-molecular domains in Ionic Liquids]
  {Nanoscale supra-molecular domains in Ionic Liquids: A statistical mechanics definition}


\begin{document}

\begin{tocentry}
Dipolar nanodroplet model of ionic liquid embedded in a generic fluid
                \includegraphics[clip=true,trim=0cm 0.1cm 0.1cm
                0cm,width=6cm]{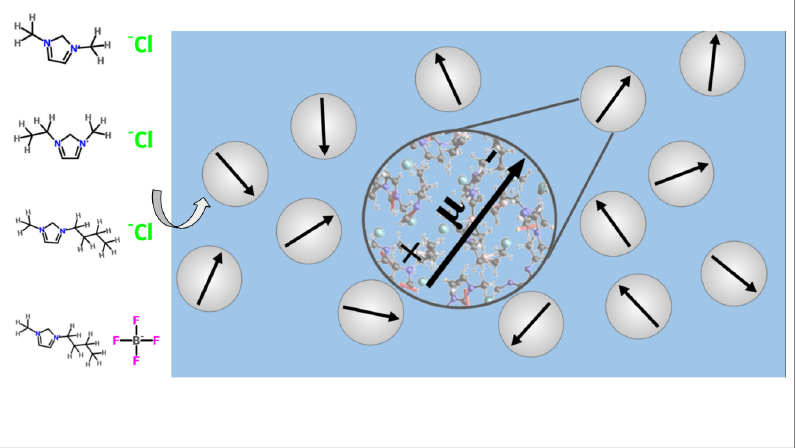}
\end{tocentry}

\begin{abstract}
 One of the many open questions concerning Ionic Liquids (ILs) is the
existence of nanoscale supra-molecular domains  which characterize the
bulk. The hypothesis of their existence does not meet a general consensus
since their definition seems based on {\it ad hoc} arbitrary criteria rather
than on general and solid first principles of physics. In this work, we propose
a definition of supra-molecular domains based on first principles of
statistical mechanics. Such principles
can be realized through the application of a recently developed  computational tool which employs adaptive molecular resolution. The method can unambiguously identify the
smallest region of a liquid for which the atomistic details are strictly
required while the exterior plays the role of a
generic structureless thermodynamic reservoir. We consider four different
imidazolium-based ILs and show that indeed one can quantitatively represent
the liquid as a collection of atomistically self-contained nanodroplets
embedded in a generic thermodynamic bath.

\end{abstract}

\begin{figure}
	        \centering
               \includegraphics[clip=true,trim=0cm 0cm 0cm
                0cm,width=8cm]{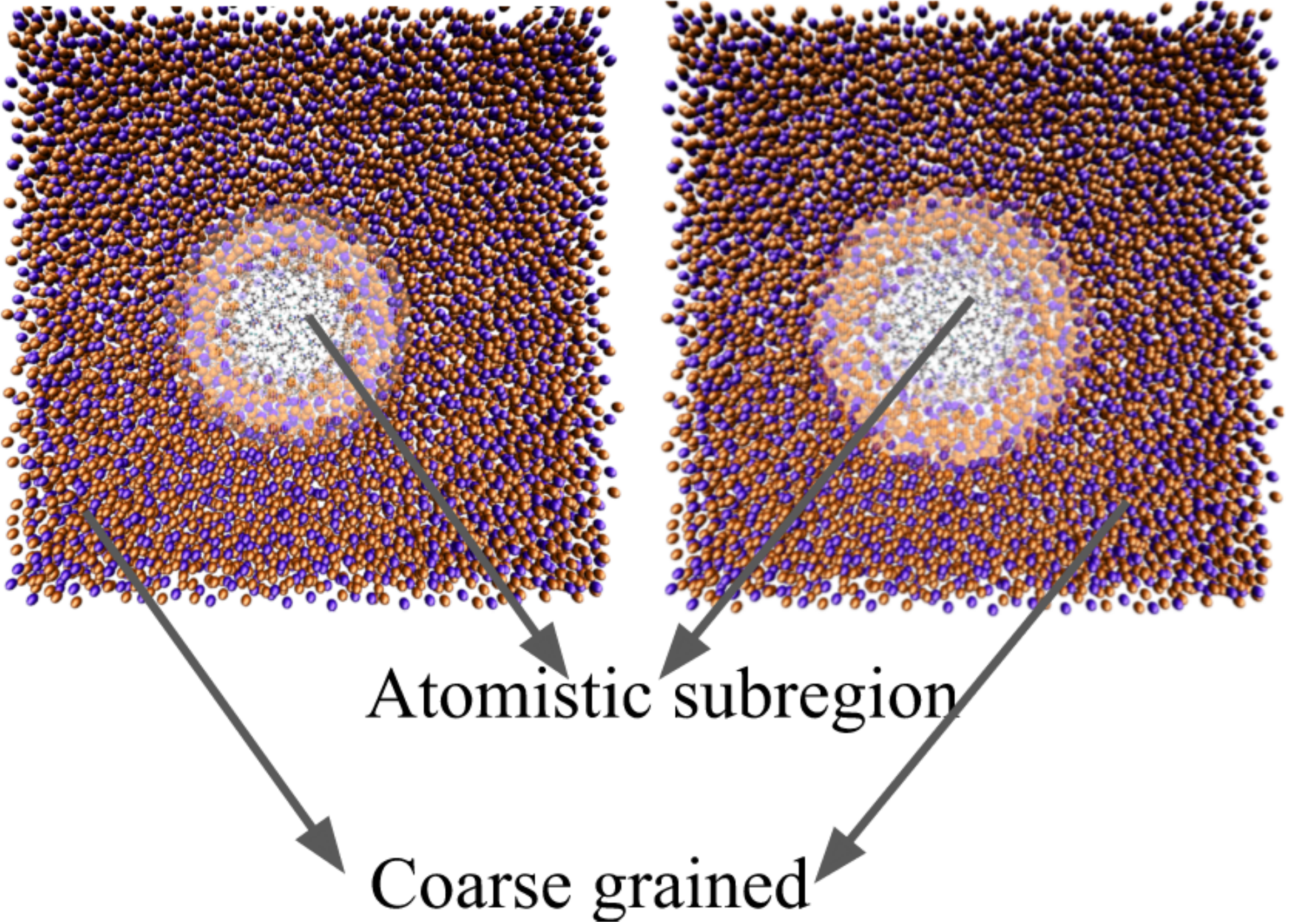}\\
               \includegraphics[clip=true,trim=0cm 0cm 0cm
                0cm,width=8cm]{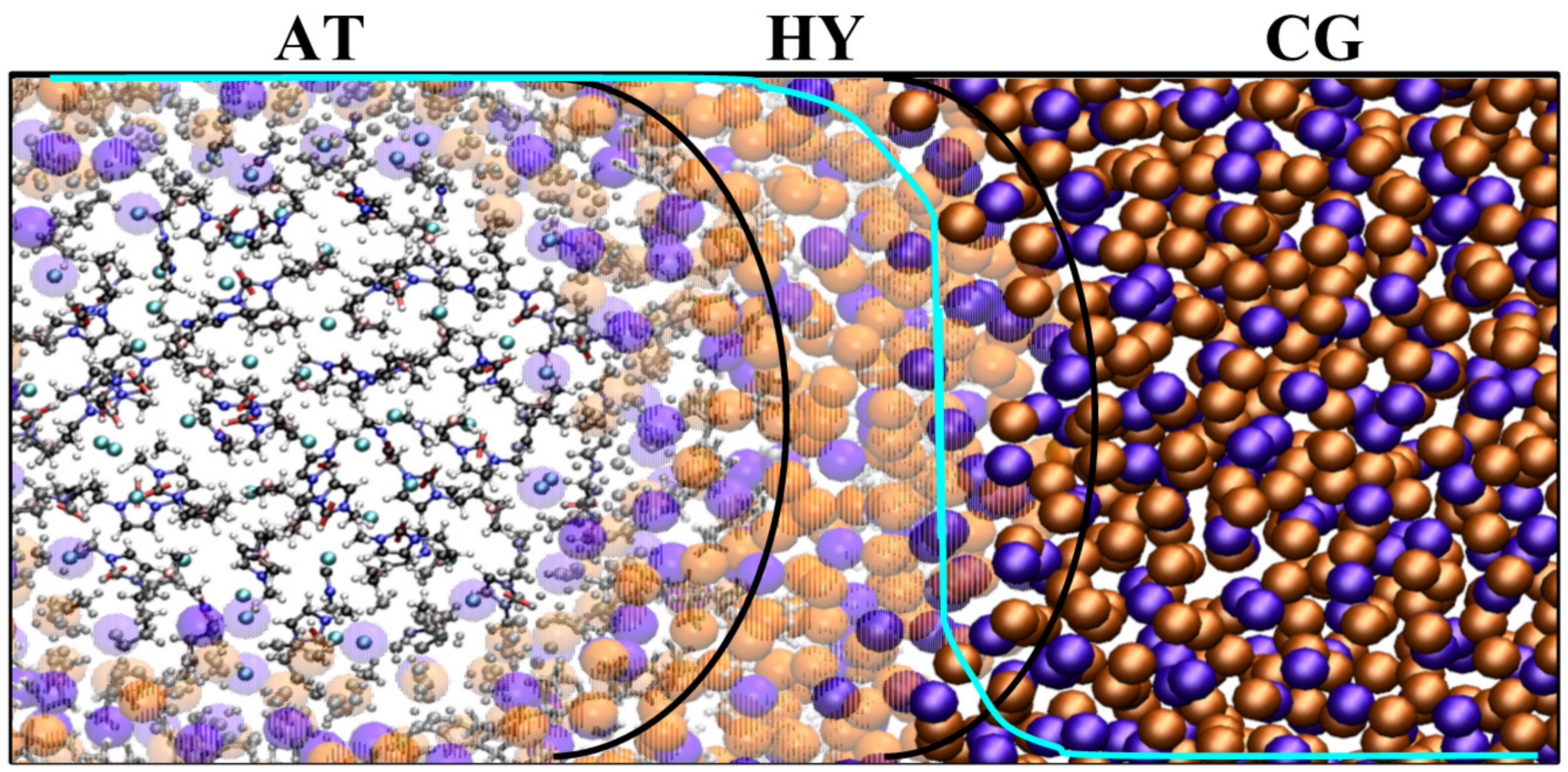}\\
               \includegraphics[clip=true,trim=0cm 0cm 0cm
                0cm,width=8cm]{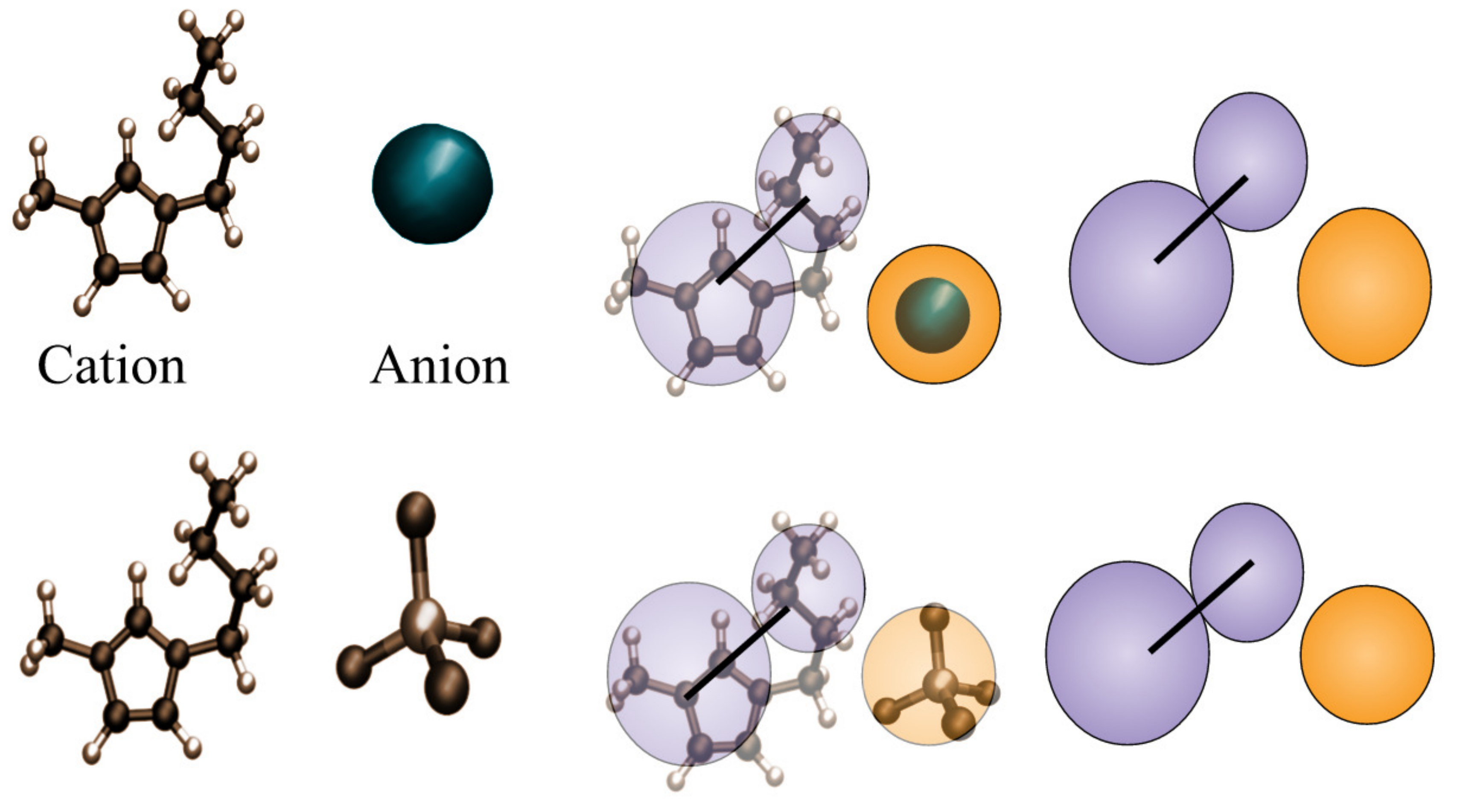}
                \caption{Graphical representation of  the GC-AdResS simulation setup.
                  (top panel) The spherical atomistic subregions for two separate simulations, 
              within which the atom-atom interactions are explicitly considered. Outside
              this region the force between two atoms is not explicitly
              treated; instead a center of mass-center of mass chargeless
              coarse-grained potential (particle-based mean field) describe
              the entire molecule-molecule interaction. (middle panel) The
              general feature of the GC-AdResS scheme is reported. 
                  Three spatial regions CG, HY and AT represents the
                  coarse-grained, hybrid and the atomistic
                  resolutions respectively. 
                The forces between the particles present in the HY region are
                derived using a smooth space dependent interpolation scheme,
                weighted by a function w(x) (cyan). Bottom panel shows the
                modeling prospect used in the AdResS. For the ILs with cationic
                alkyl chain length less than 1 (i.e.,) for [DMIM][Cl], we use
                one coarse grained bead  to model the cation, and for the other
                systems we have used two CG beads to model the entire cation. In
                the same way we have used one CG bead to  model the 
              spherical anions.}
                \label{fig:cartoon.il}
                    \end{figure}

\begin{figure}[!htbp]
	        \centering
               \includegraphics[clip=true,trim=0cm 0cm 0cm
               0cm,width=14cm]{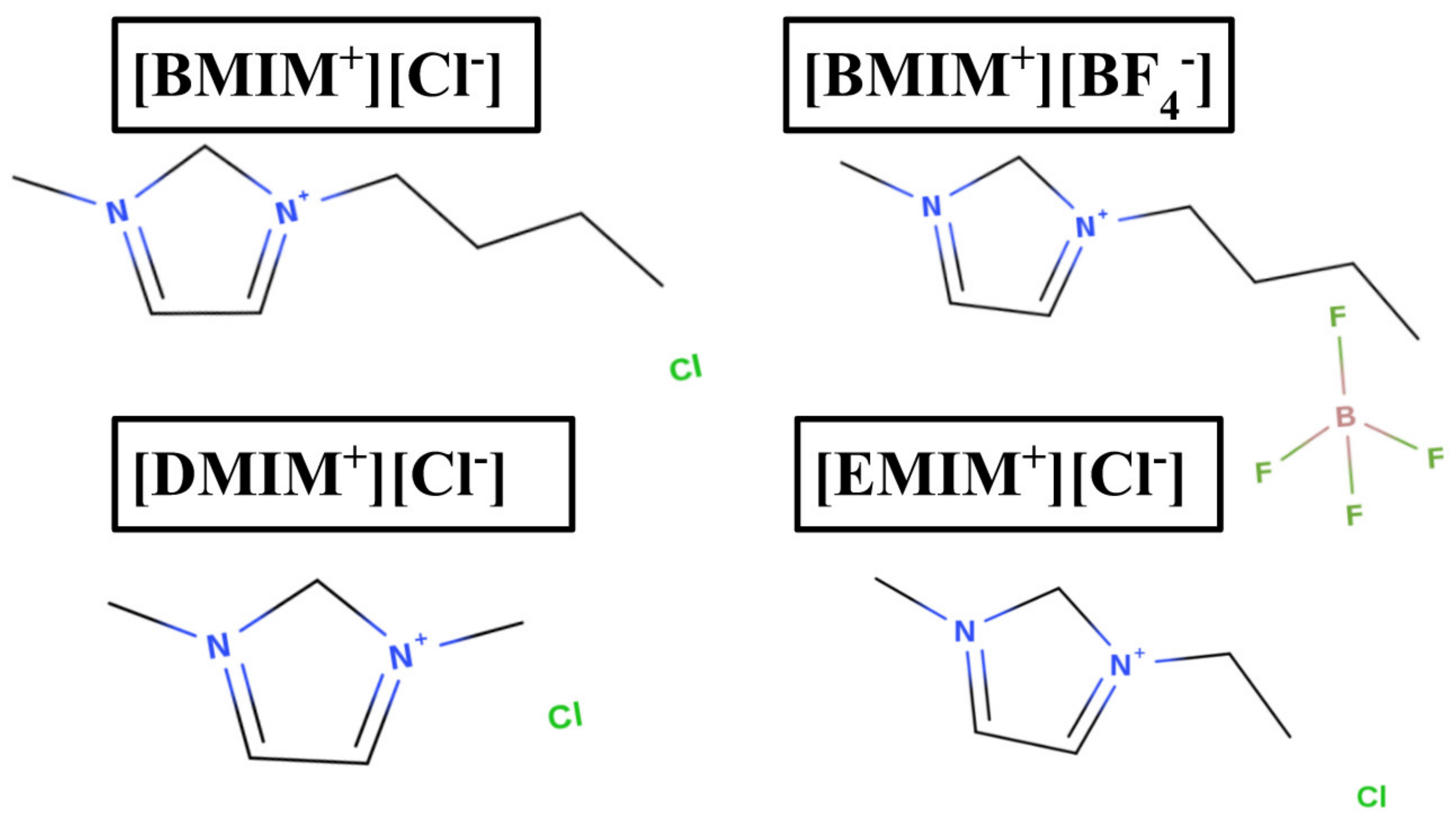}
                \caption{Chemical structure of the ILs used in the study}
                \label{fig:chemicalstructure.il}
                \end{figure}

Ionic liquids (ILs) are an unusual class of liquids where the complex interplay of van der Waals, electrostatic and short-ranged solvophobic
interactions determines the formation of 
alternating nano-metric structures in the bulk or at interfaces.
The non-polar alkyl chains form non-polar
domains by excluding themselves solvophobically from the ionic polar domains, in turn,  this ability to spontaneously form self-assembled nano-structures without the need of external factors, allows them to play a key role in the evolution of modern nanotechnology.
The list of properties of ILs which are used in current (nano)-technology is very broad, interested readers can consult the referenced literature \cite{lllzc08,znzqzhl10,fki03,lcwhj14,sn13,gfb14,gsp09,ur04,wv05,cp06,wv06,trbd07}.
Some relevant examples regarding Imidazolium based ILs, i.e. the class of systems considered in this work,  are medicinal applications for the treatment of gram-negative and gram-positive bacteria's, fungi,
algae, human tumor cell lines\cite{bce10,km09}.
Importantly, it turns out that all these technological applications are often {\it limited} by the stability or dimensions of the nanoscale domains in bulk or at the interface.
In fact, it has been suggested that the optimization of macroscopic properties
such as heat capacity and compressibility is directly linked to the specific
structure of such nanodomains. The natural consequence is that a rational design
of the specific molecular chemistry for the optimization of a macroscopic property needs the knowledge about the ability of chemistry-specific ion
pairs in building nanodomains with certain characteristics \cite{az13,hwa15}.\\
As a consequence, the key question that chemical engineers must face concerns
the determination of the {\it optimum} number of ions  that account for the
formation of stable apolar/polar nanoscale domains according to the specific
chemical structure of the ions. The question can be actually reverted by asking
how to specifically design a molecule (ion pair) so that a desired size of the
nanoscale domain is achieved. Molecular design via computer simulation can play
a major role in addressing such a question, in fact the chemical structure of a
molecule can be modified {\it in silico} and then its liquid properties can be
simulated at the thermodynamic conditions desired. The success of molecular
simulation in explaining the important properties of ionic liquids and in
promoting computational design of new species has been proven by a large
number of publications (see
Refs.\cite{wv05,wv06,ur04,cp06,wzdbhkd11} and references therein).
In literature, the definition of nanodomains in ILs is analyzed with a severe
critical eye and some of the related results, such as the critical number of
ions, are considered ``magic'' data, because associated to an arbitrary
length scale which does not have proven physical or chemical basis \cite{gsacde04,hwa15}. In general one can define nanoscale domains such domains that arise due to solvophobic self-assembly and arising because of the amphiphilic nature of the ions; these can be thought of having an experimental meaning since their characteristic size is measured by x-ray scattering. In addition, there are small regions or aggregates which behave as if they hold one unit of charge (and thus giving a low ion concentration overall). This is usually considered only a physical model which is used to explain the observed screening length; these aggregates are not directly measured (or measurable).
In this context, this work aims at contributing with an original analysis of
possible nanodomains for a class of imidazolium-based ILs. Specifically, we aim
at identifying the smallest region of the liquid whose properties strictly
require the explicit consideration of atomistic degrees of freedom. In relation to the discussion above, both kinds of aggregates, i.e. those produced by the mechanism of solvophobic self-assembly and those which behave as a unit of charge, require (locally in space) the explicit chemical structure of the molecules. The extension in space for which the explicit consideration of the atomistic degrees of freedom (and thus charges) is mandatory would give information on the interplay between the two type of aggregates in terms of the interplay between entropy (molecular packing) and energy (direct atom-atom interactions). Such an interplay, in turn, can lead to a physical model of nanodomain that contains both kinds of aggregates, each with its specific signature within the definition of a nanodomain given by our analysis. It must be added that not all ILs are characterized by the heterogeneity of the first kind, only ILs with alkyl chain length $n\ge 4$ \cite{hwa15}. For this reason we will consider ILs with alkyl chain length up to 4, so that we can treat both kinds of aggregates. The
proposed analysis implies that in a subregion of the system the explicit chemical
structure of the molecules is necessary for any calculated property, instead the
rest of the system plays the role of a generic thermodynamic environment where
explicit chemical details do not play any role.\\
Such an analysis is possible with a simulation tool developed in our group over
the last few years, that is the Grand Canonical Adaptive Resolution Simulation
method (GC-AdResS)
\cite{jcp2010-simon,jcp-2015-pi,jcp-2010-full,annurev2008,whss13,azhws15}. Such
a method allows for treating open subregions of a system at different level of
resolution, specifically a high resolution region (atomistic) embedded in a
large uncharged coarse-grained mean-field environment (see
Fig\ref{fig:cartoon.il}).
The method is based on a rigorous numerical implementation of Grand Canonical
principles (for further details see Refs. \cite{prl2012,lds16} and references therein); a relevant consequence for the current work is that the smallest atomistic region in GC-AdResS that can reproduce certain quantities of a full atomistic simulation, unambiguously defines a self-contained atomistic (sub)region (atomistic nanodroplet, w.r.t such quantities). In this perspective we will consider a series of imidazolium-based ILs, 1,3-dimethylimidazolium
chloride (DMIM Cl), 1-ethyl-3-methylimidazolium chloride, (EMIM Cl), 1-butyl-3-methylimidazolium
 chloride (BMIM Cl) and 1-butyl-3-methylimidazolium tetrafluoroborate (BMIM
 BF$_4$) (see Fig.\ref{fig:chemicalstructure.il}) and apply the GC-AdResS analysis to define the nanodroplet domain and analyze its physical and chemical characteristics.
 Using state of the art force fields
 available in literature \cite{dwbsh12,lrsk10}, we will show that one can
 quantitatively describe the liquid as a collection of supra-molecular domains
 embedded in a chargeless particle-based mean-field. As a matter of fact, we
 provide an objective route to the definition of nanodomains with their
 associated physical properties. A relevant consequence of our analysis is the
 direct link between the
 microscopic scale of specific chemistry and mesoscopic structure of the
 liquid; in perspective, this link offers a possible route to a systematic rational design of large scale liquid properties.
\subsection{ Determination of the size of nanodroplets via GC-AdResS analysis}
In order to determine the size of the atomistic nanodroplets we apply the
GC-AdResS tool to the liquid. The application of such a tool implies that
quantities calculated in the atomistically resolved region carry the full chemical
characteristics of the molecules of such a region, while the influence of the
exterior (coarse-grained region) enters only in the form of an action of a generic
bath that thermodynamically equilibrates the atomistic region. The implication is that if a certain quantity
is calculated in the atomistic region of GC-AdResS and displays the same
behavior when calculated in the
equivalent subregion of a full atomistic simulation, then the explicit role of
atomistic degrees of freedom outside such region is negligible for the quantity
considered. As a consequence, with respect to such a quantity, the atomistic
region of GC-AdResS can be considered as an atomistically self-contained
nanodroplet embedded in a liquid of uncharged  coarse-grained spheres( one for
the anion and one, or more bonded, for the cation). In essence, the
exterior represents a generic fluid, i.e. without any information regarding
the chemical structure of the molecules, that operates
as a large thermodynamic bath for the atomistically resolved droplet. The quantification of the minimal size of
such subdomain is determined by running a sequence of GC-AdResS simulations
with decreasing size of the atomistic region. The smallest atomistic region
for which the target property has the same behavior as in a full atomistic
simulation (within maximum $5\%$ of discrepancy) defines the size of the nanodroplet. The choice of the quantity to use as a target for the
definition of an atomistic nanodroplet plays a key role; it must be such that on the one hand it
carries a specific signature of the system and on the other it should be a
quantity that can be directly used to determine a broad range of physical properties. In previous work
\cite{whss13} we have proposed that atomistic radial distribution functions are
ideal quantities to use as target. In fact they uniquely characterize the liquid
structure of a substance and in addition, at statistical mechanics level, the
probability distribution function of an atomistic system can be approximated as
a factorization of (all possible) atom-atom radial distribution functions
(two-body correlation functions): 
\begin{equation}
P({\bf r}_{1}........{\bf r}_{N})\approx
\Pi_{i,j}g(r_{ij})
\label{distfun}
\end{equation}
where ${\bf r}_{1}........{\bf r}_{N}$ is the spatial configuration of the $N$
molecules of the liquid, $i$ and $j$ labels atoms belonging to
different molecules at a distance $r_{ij}$. Since the knowledge of $P({\bf r}_{1}........{\bf r}_{N})$
implies the knowledge of every average property/quantity of the system, its
approximation via the factorization proposed above implies the complete
statistical knowledge of the system up to the two-body level (second order). The determination of the size of
the nanodroplets using GC-AdResS is done by determining the smallest atomistic
region for which {\bf every possible} atom-atom radial distribution function
has the same behavior as the corresponding quantity calculated in the
equivalent subregion of a full atomistic simulation. In addition, being the
atomistic region an open system governed by Grand Canonical principles, it is
mandatory to require that the
molecular number probability, $P(N)$,  must have the same behavior (within a
certain accuracy) of the , $P(N)$, calculated in the equivalent subregion
of a full atomistic simulation of reference. In molecular simulation, for a system at uniform resolution (i.e. full atomistic), it is trivial to obtain all the atomistic radial distribution functions, provided that the size is sufficiently large to justify the statistical average. In our specific case the situation is very different: outside the atomistic region the atomistic structure of the liquid does not exist, thus it is not obvious to reproduce the atom-atom radial distribution functions in the atomistic region of AdResS. The agreement is possible only if the atomistic information of the bulk is not required for the local structure-based microscopic properties. We must clarify that here the dynamics is not treated, and that we focus only on static properties.
 This criterion has
been proven to be unambiguous and effective in identifying relevant length scale
at nanoscopic level in previous studies of relevant systems
\cite{jcp-2010-full,pccp-2017-full,jkd18}, and in preliminary work on ILs
\cite{jkkd18,jk18,ks17}. At this point
a warning must be added, that is, Eq.\ref{distfun} implicitly means that properties
where three-body (and higher) accuracy is required cannot be
treated with the current approach, unless the corresponding n-th-body
correlation function is used as a target. This means that the extension
to higher accuracy and further statistical properties, is technically
possible\\
In the next section we report the
calculations done for each system and we quantitatively define the nano-metric
characteristics of each liquid. Furthermore, as a relevant physical property
of such nanodomains, we calculate the total dipole
moment of the atomistic region of GC-AdResS and compare it with the equivalent
quantity calculated in the same sub-domain of a full atomistic simulation and
prove that indeed the values agree within a maximum discrepancy of
$5\%$. This calculation is of particular interest, in fact implies the
existence, within the defined nanodomains,
of smaller domains with a predominance of
positive and negative ions. 
\section{Results}
\begin{figure}[!htbp]
	        \centering
                \includegraphics[clip=true,trim=0.cm 0.cm 0.cm
                0cm,width=15cm]{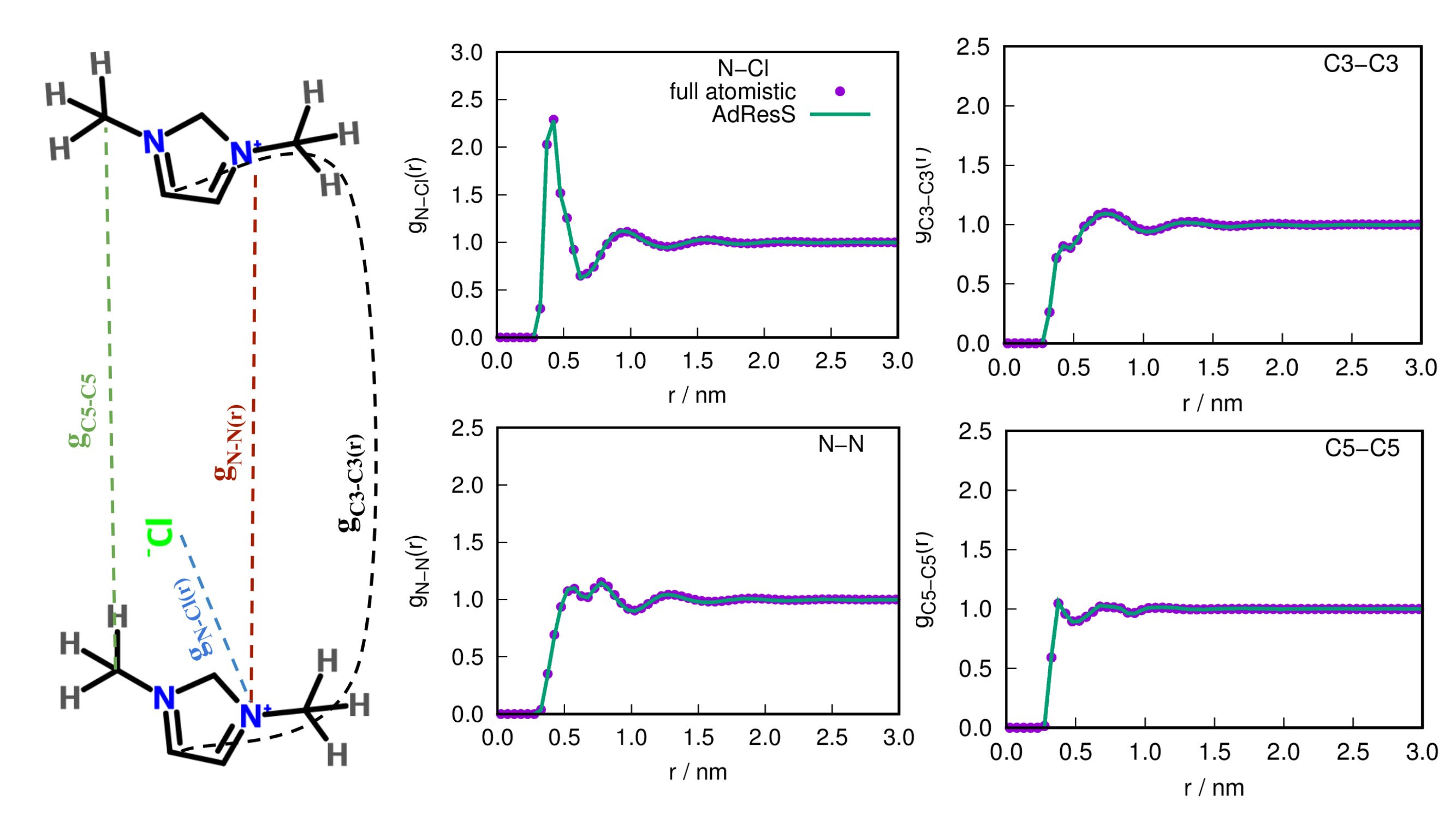}\\
                \caption{Three example of atom-atom radial distribution function
                of [DMIM][Cl] calculated in AdResS and in the equivalent subregion of a full atomistic simulation. Here are reported only results of the simulation with the smallest atomistic region of AdResS possible. In simulations with a smaller atomistic region, the radial distribution function show a significant deviation from the results of a full atomistic simulation. Here, C3 is carbon atoms in the 5th position of the imidazole ring. C5, C6, C8 are the tail carbon atom of the alkyl chain attached to the 3rd position of the imidazole ring.}
                \label{fig:rdfmmimcl}
                \end{figure}

\begin{figure}[!htbp]
	        \centering
                \includegraphics[clip=true,trim=0.cm 0.cm 0.cm
               0cm,width=15cm]{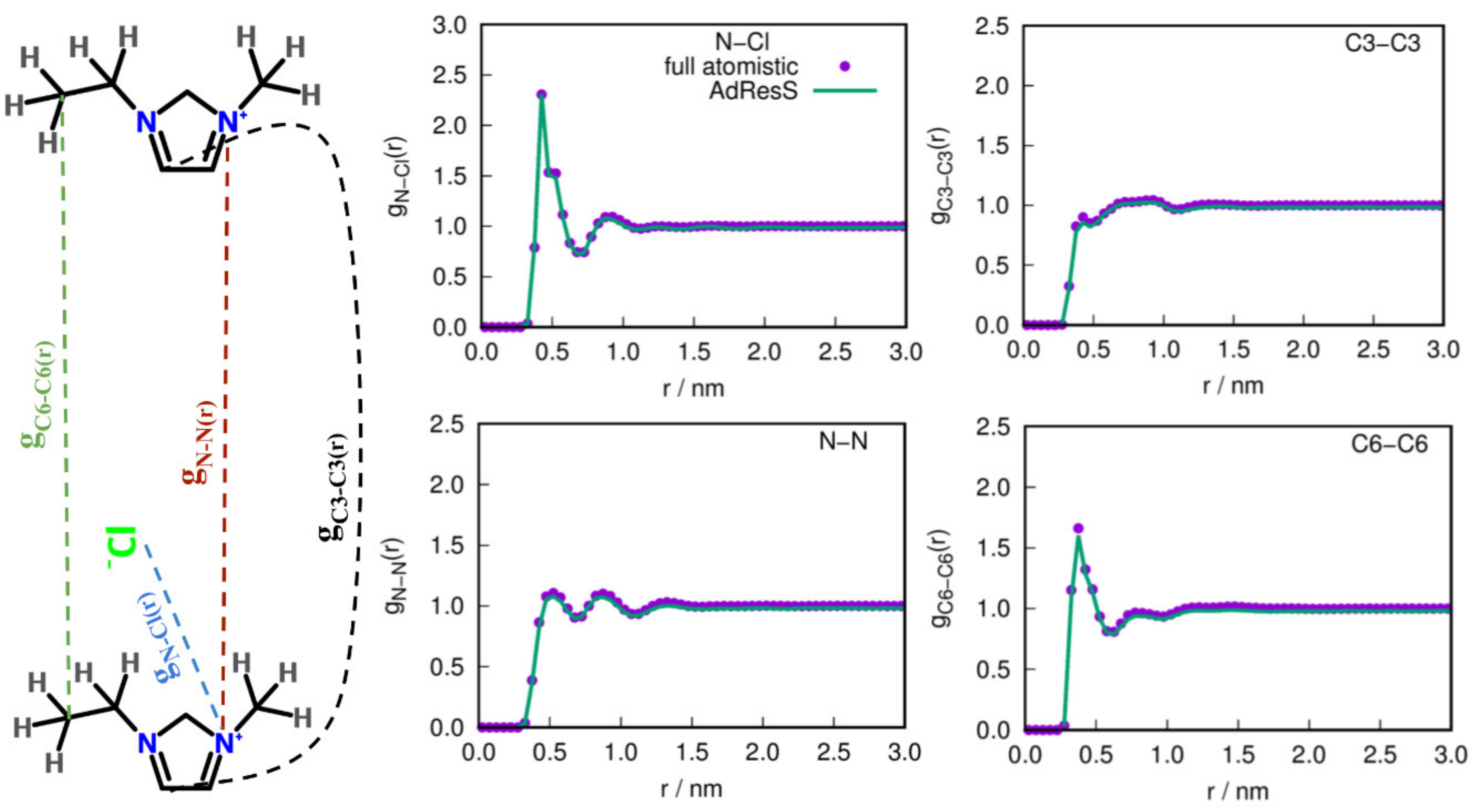}\\
                \caption{As in the previous figure for [EMIM][Cl].}
               \label{fig:rdfemimcl}
                \end{figure}

\begin{figure}[!htbp]
	        \centering
                \includegraphics[clip=true,trim=0.cm 0.cm 0.cm
                0cm,width=15cm]{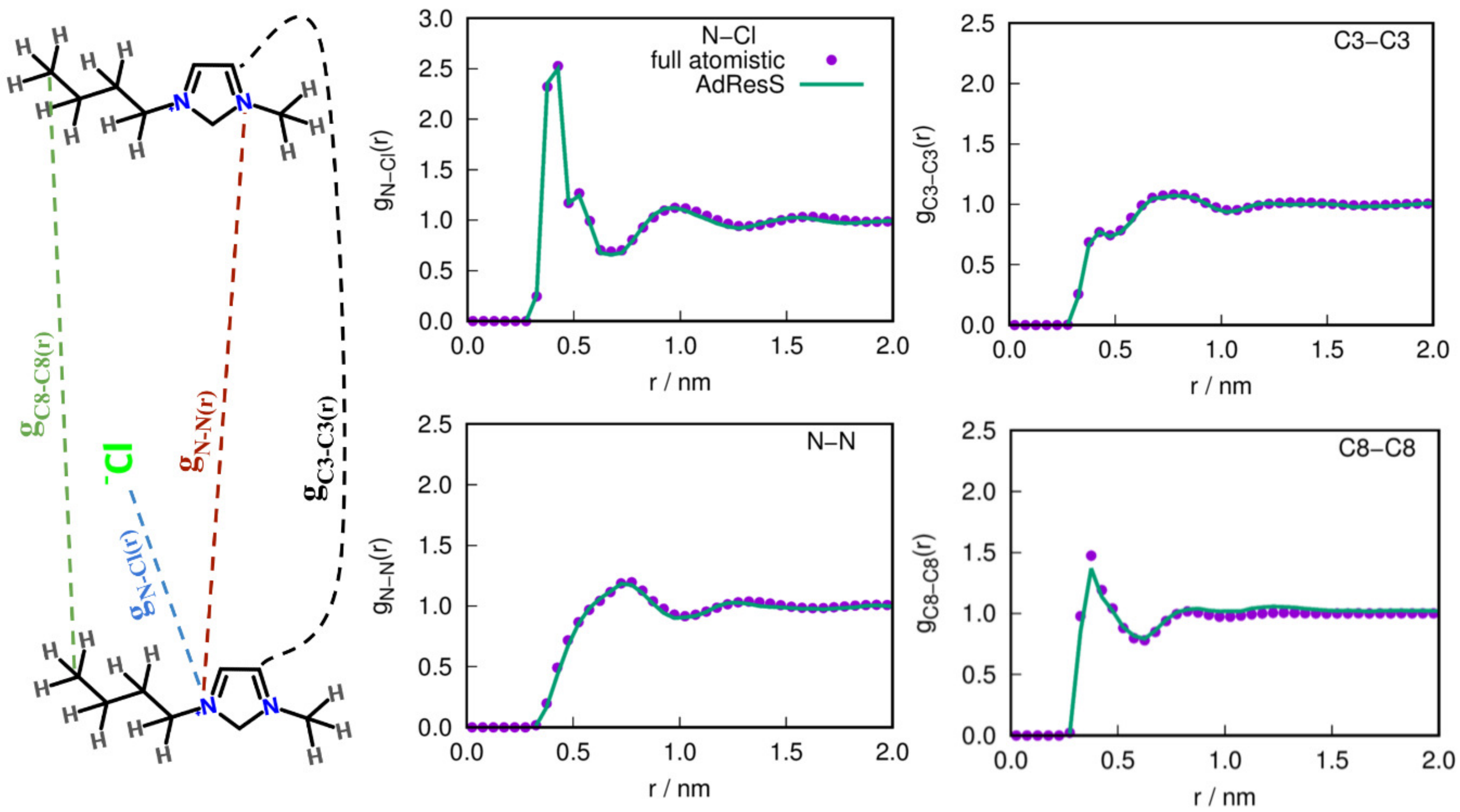}\\
                \caption{As in the previous figure for [BMIM][Cl].}
                \label{fig:rdfbmimcl}
                \end{figure}

\begin{figure}[!htbp]
	        \centering
                \includegraphics[clip=true,trim=0.cm 0.cm 0.cm
                0cm,width=15cm]{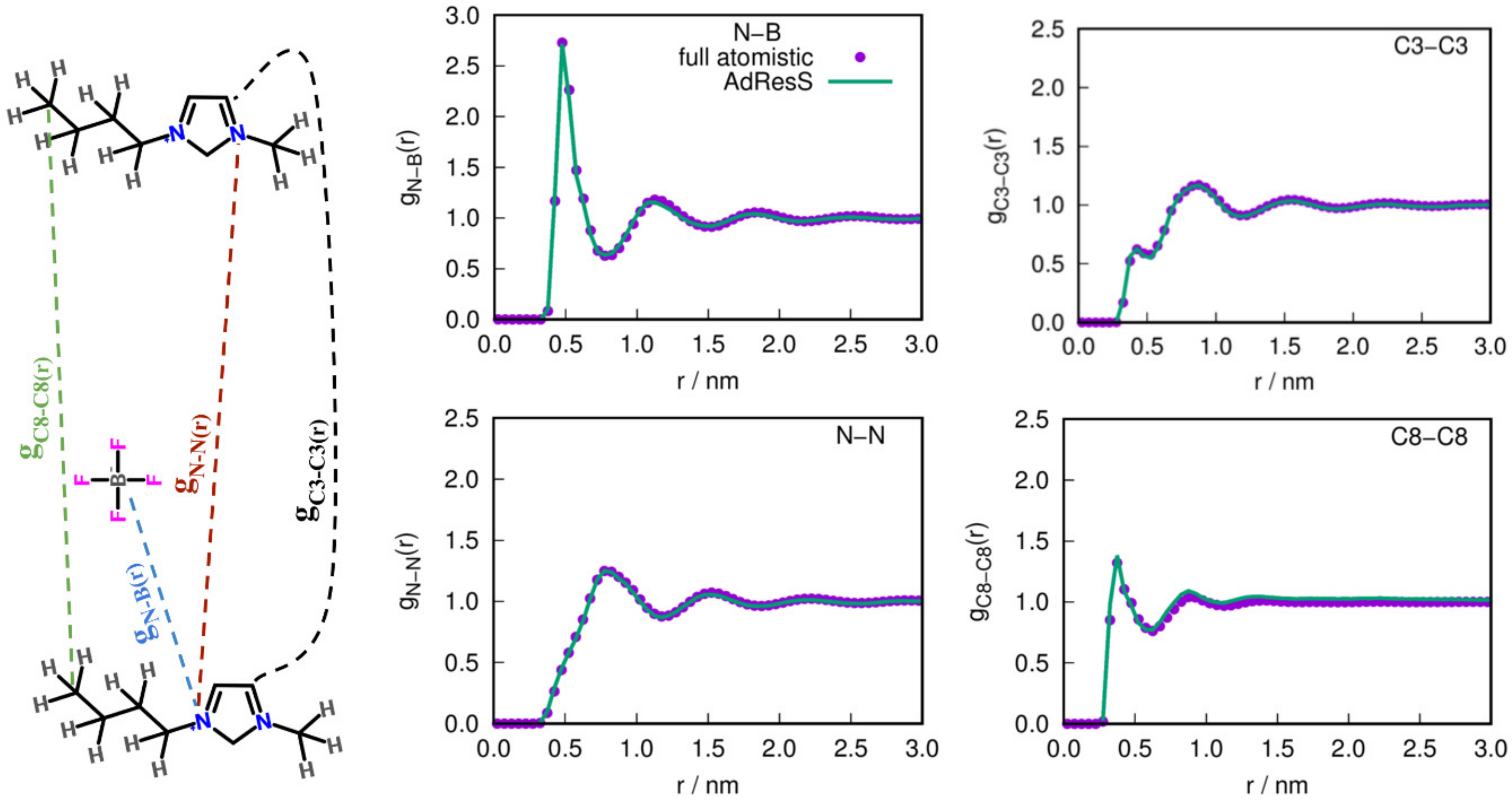}\\
                \caption{As in the previous figure for [BMIM][BF$_4$].}
                \label{fig:rdfbmimbf4}
                \end{figure}

The technical details of the simulation are reported in the methods section (for further details, see \cite{jkkd18,jk18,ks17}). Here we report the results of the
atom-atom radial distribution functions
and molecular number probability.  It must be underlined once again that we consider {\bf all possible
atom-atom radial distribution functions} (according to the symmetry of
the molecules), however for clarity of presentation and in order to avoid an excess of
equivalent information
Figs.(\ref{fig:rdfmmimcl}-\ref{fig:rdfbmimbf4}) report only some
representative examples of
the atom-atom radial distribution functions calculated in GC-AdResS and
compared with the same quantities calculated in the equivalent subregion of a
full atomistic simulation.
All the other functions
are not reported explicitly, but they are characterized by the same accuracy
regarding the agreement between the AdResS simulations and the full atomistic
simulation of reference.
The size of the atomistic domain varies between $2.0$
and $3.0$ nm for the different ILs considered (see table \ref{tab:locality.charge}; as made clear before, outside the atom-atom radial distribution functions 
do not exist, thus if the atomistic data of the exterior were important for the
radial distributions in the atomistic region, then one must have seen a clear perturbation
of the functions. 
\begin{table}
\caption{Radius of the minimal atomistic region of AdResS (size of the nanodroplet) for each system considered.}
\begin{tabular}{ccc}
\hline
System &  Radius (nm) \\
       
\hline
$[DMIM][Cl]$ &  3.0 \\ 
$[EMIM][Cl]$ &  3.0 \\
$[BMIM][Cl]$ &  2.0 \\
$[BMIM][BF_4]$ & 3.0 \\
\hline
\end{tabular}
\label{tab:locality.charge}
\end{table}
As anticipated above, the atomistic region of GC-AdResS is an open
system, thus the definition of a statistically self-contained atomistic region
would be justified if and only if the exchange of particles with the exterior
occurs in the same way as in an equivalent subregion of a full atomistic
simulation. Such a behavior is expressed by $P(N)$ i.e. the particle number
probability, 
\begin{figure}[!htbp]
                \centering
                \includegraphics[clip=true,trim=2cm 2cm 6cm
                16cm,width=8cm]{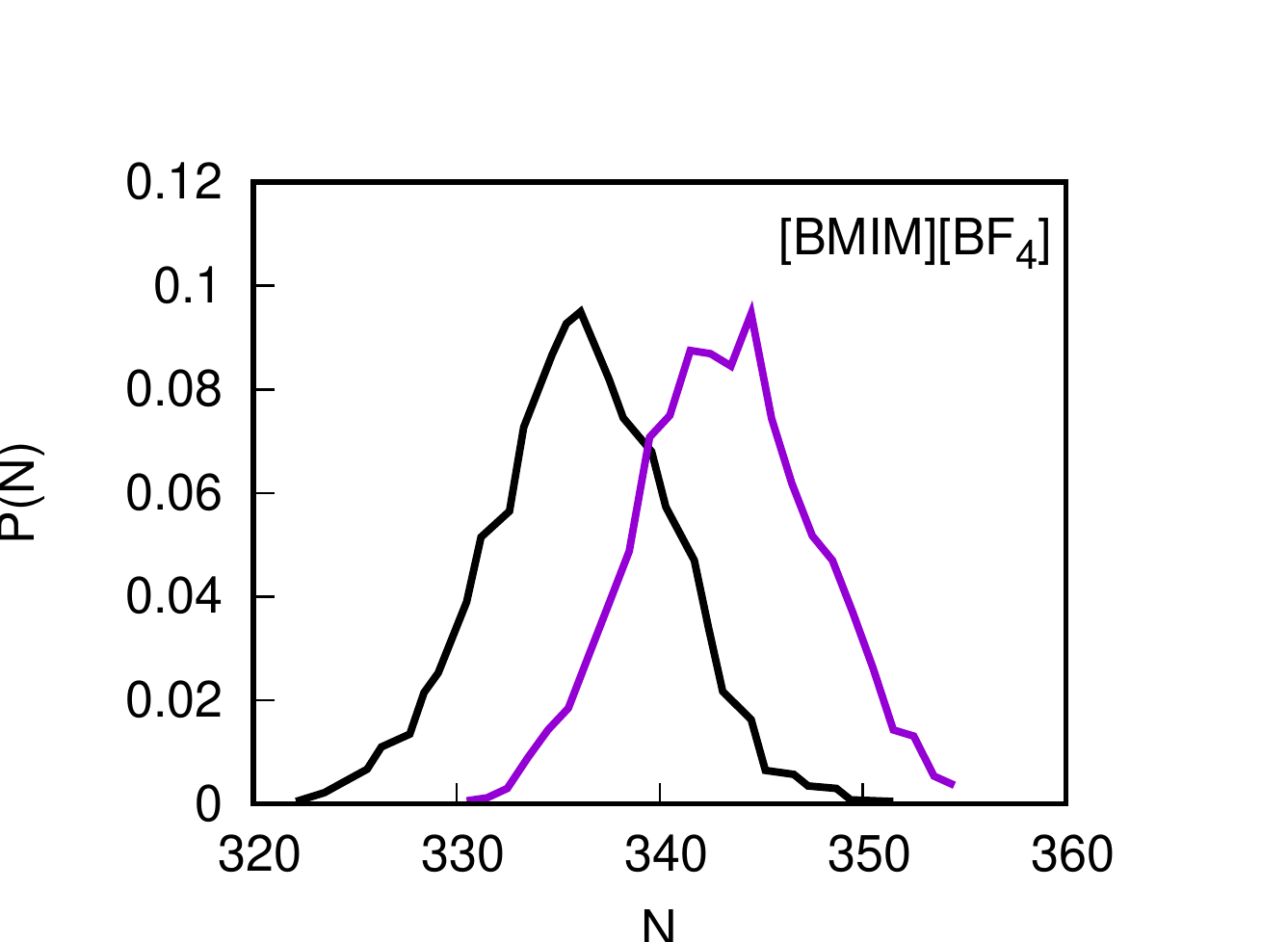}
                \includegraphics[clip=true,trim=2cm 2cm 6cm
                16cm,width=8cm]{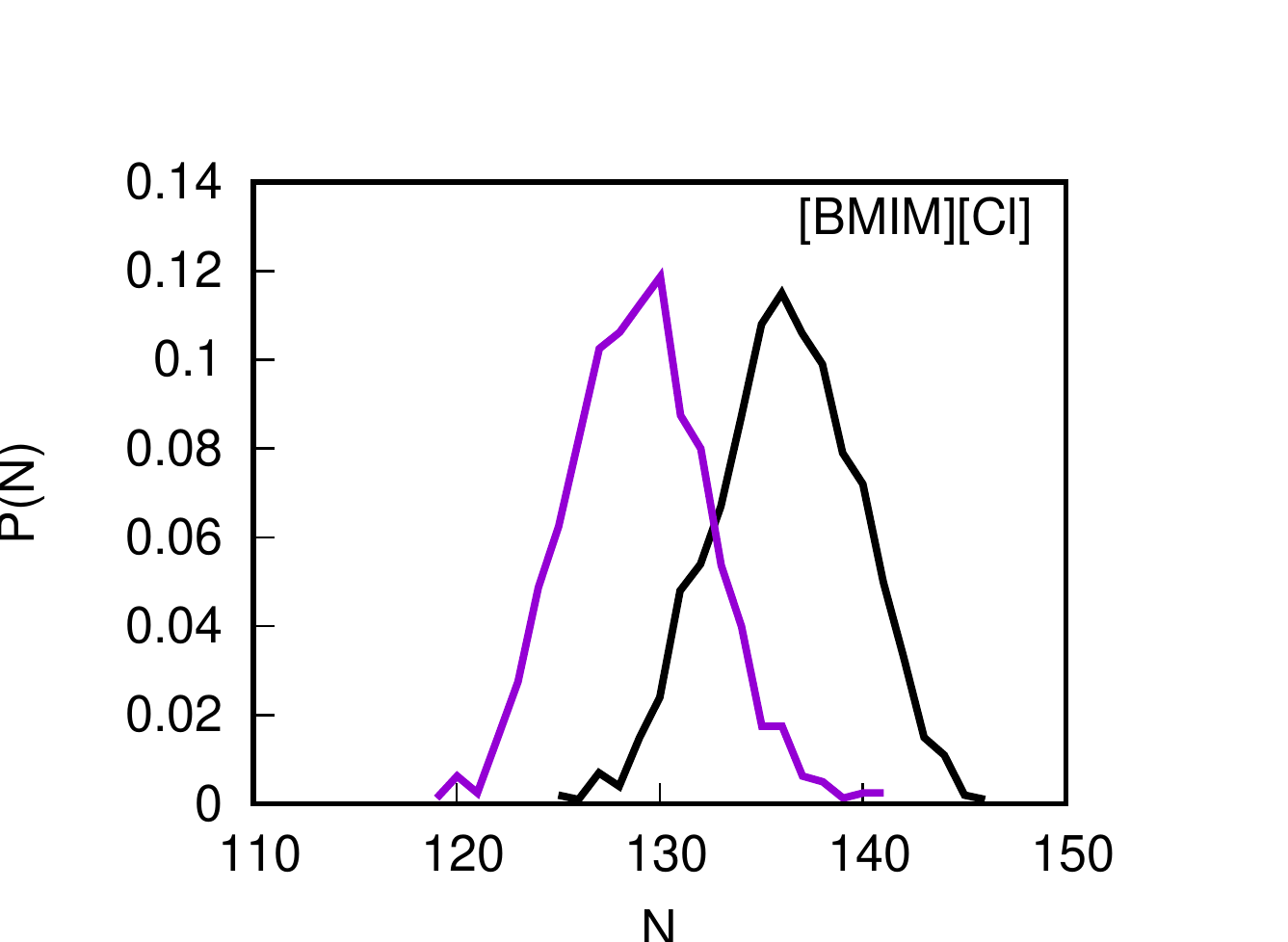}\\
                \includegraphics[clip=true,trim=2cm 2cm 6cm
                16cm,width=8cm]{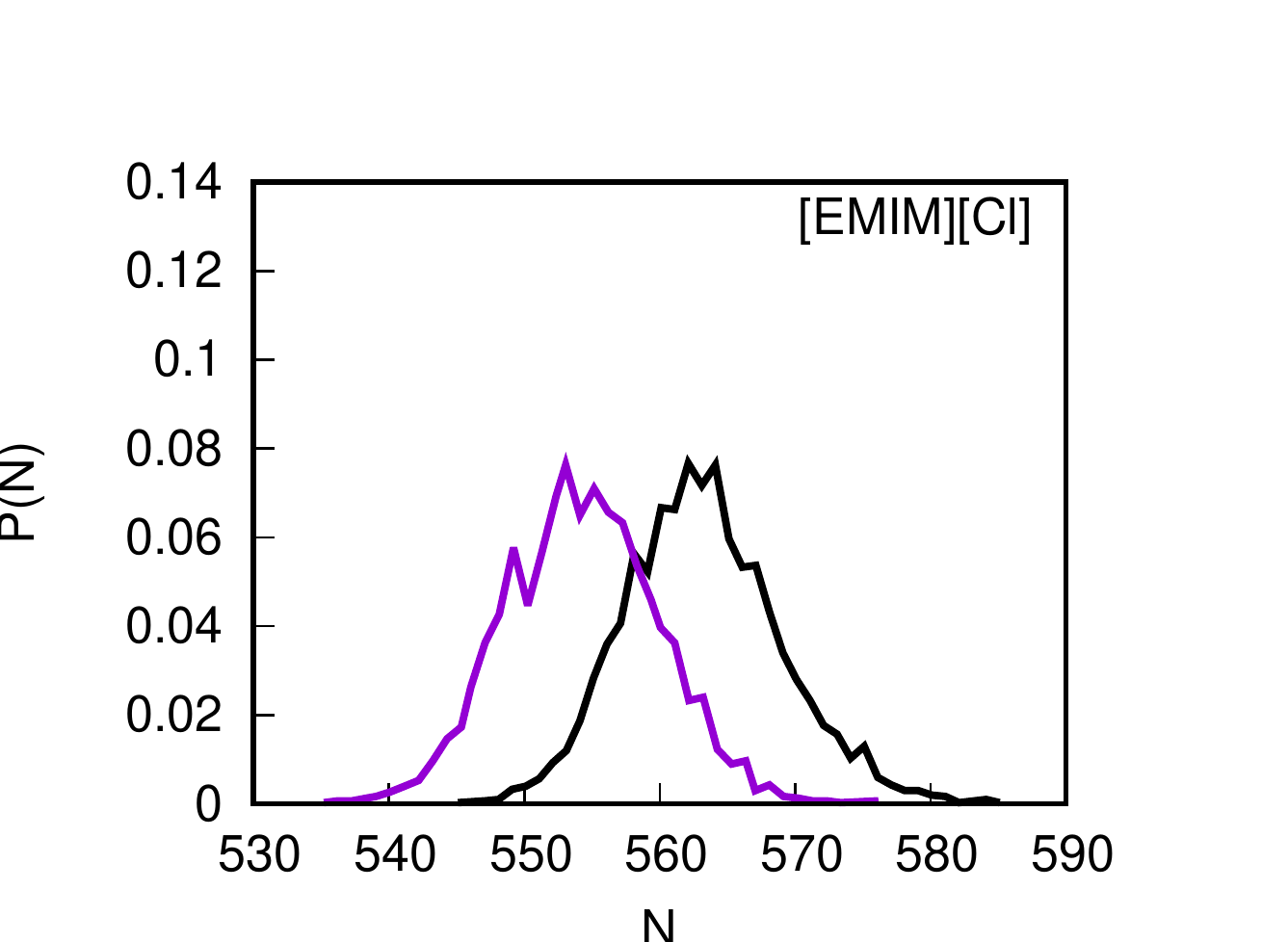}
                \includegraphics[clip=true,trim=2cm 2cm 6cm
                16cm,width=8cm]{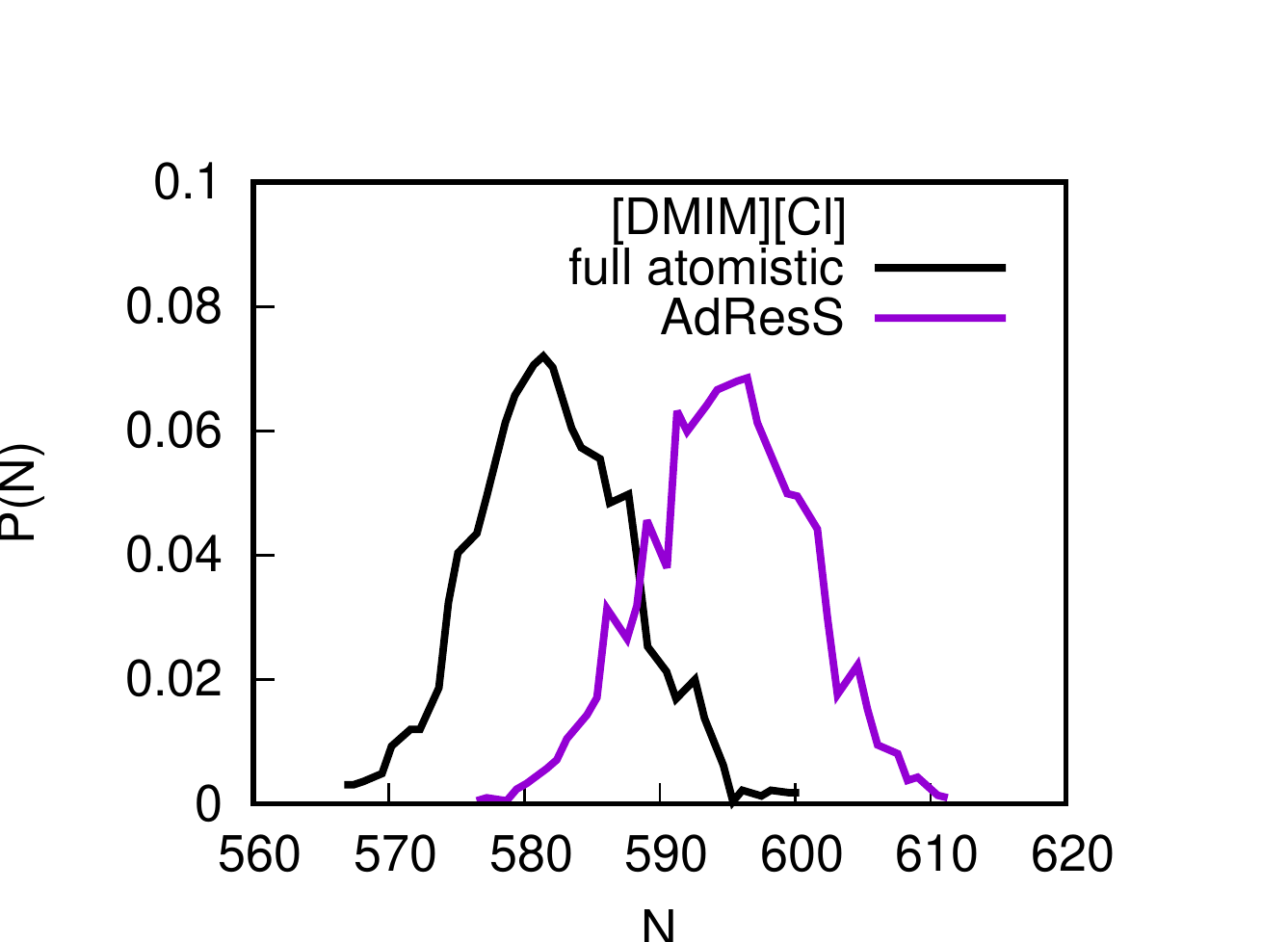}
                \caption{Probability distribution of the average number of ion
                  pairs computed in the atomistic region of the AdResS and
                  compared with the same quantity computed in the equivalent
		subdomain of the full atomistic simulations for the ILs (see table \ref{tab:avg.pn}).}
                \label{fig:POFN}
                \end{figure}
Fig.\ref{fig:POFN} shows that for every system considered the
atomistic domain identified with GC-AdResS display a very good agreement with
the full atomistic simulation of reference regarding $P(N)$ (within $3\%$ accuracy).
In the light of the results reported above, one can be certain that the representation of the exact many-body probability distribution
function of the atomistic domain is correct up to the second order. As a
consequence any quantity calculated as an ensemble average in the atomistic
region has the same value when calculated in the equivalent subregion of
a full atomistic simulation of reference, i.e. we can identify local
characteristics of the liquid in space by considering only one atomistic
droplet defined according to the criteria listed above.  In this perspective, our
proposed definition of a nanodomain can be considered very solid from the point of view of statistical mechanics.
\subsection{Dipole Moment of the Nanodroplet}
\begin{figure}[!htbp]
	        \centering
                \includegraphics[clip=true,trim=2cm 2cm 6cm
                16cm,width=8cm]{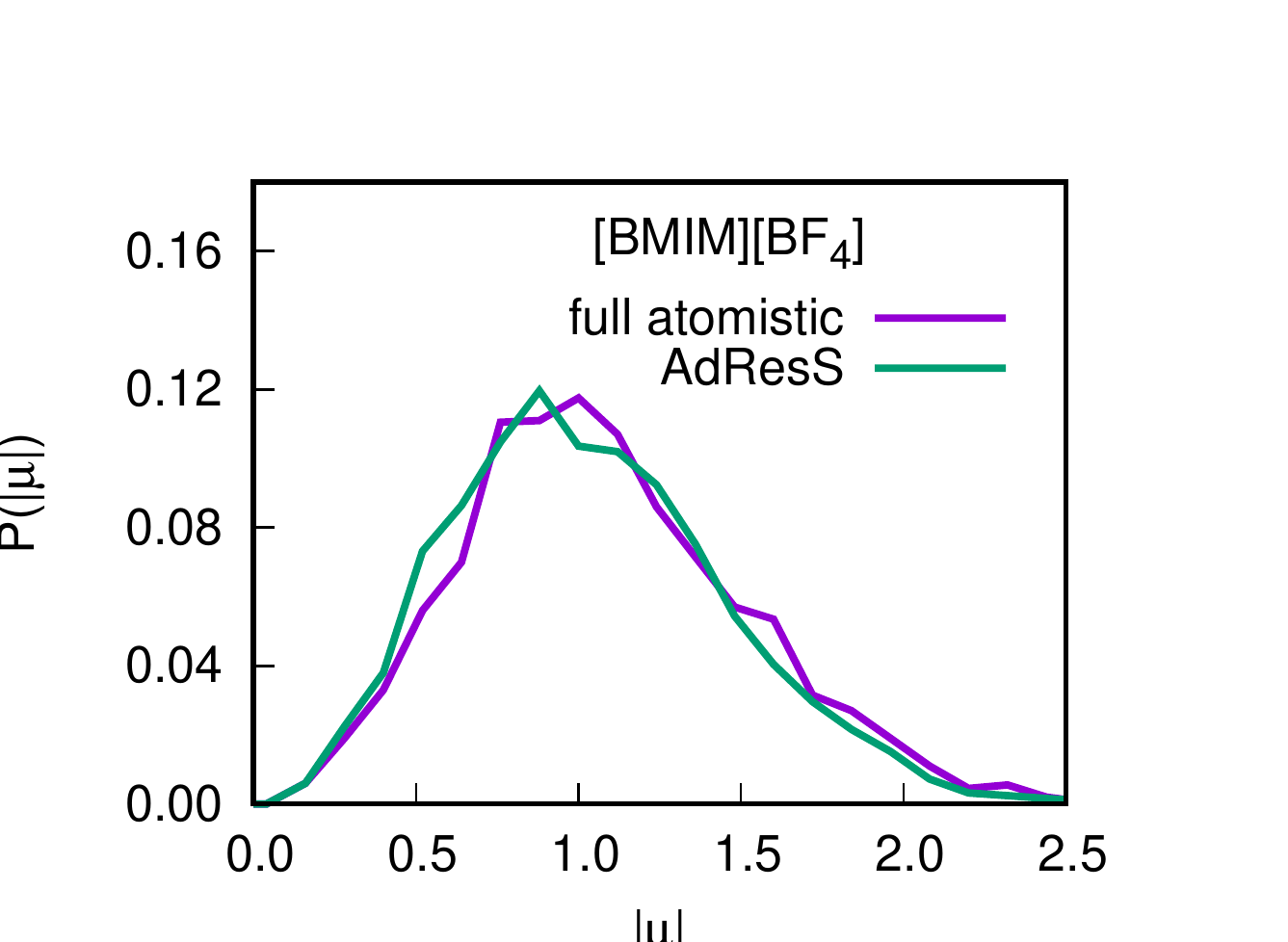}
                \includegraphics[clip=true,trim=2cm 2cm 6cm
                16cm,width=8cm]{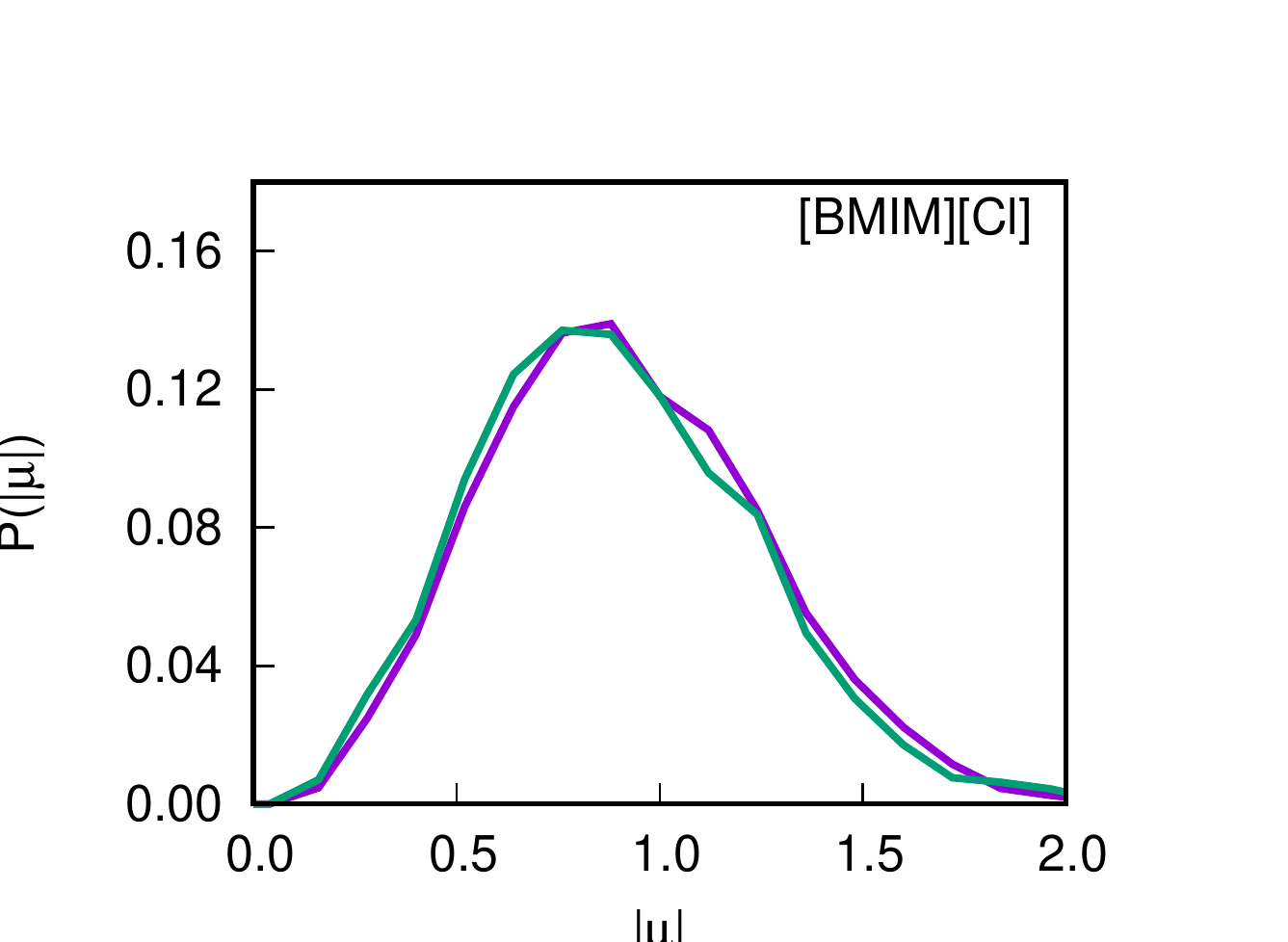}\\
                \includegraphics[clip=true,trim=2cm 2cm 6cm
                16cm,width=8cm]{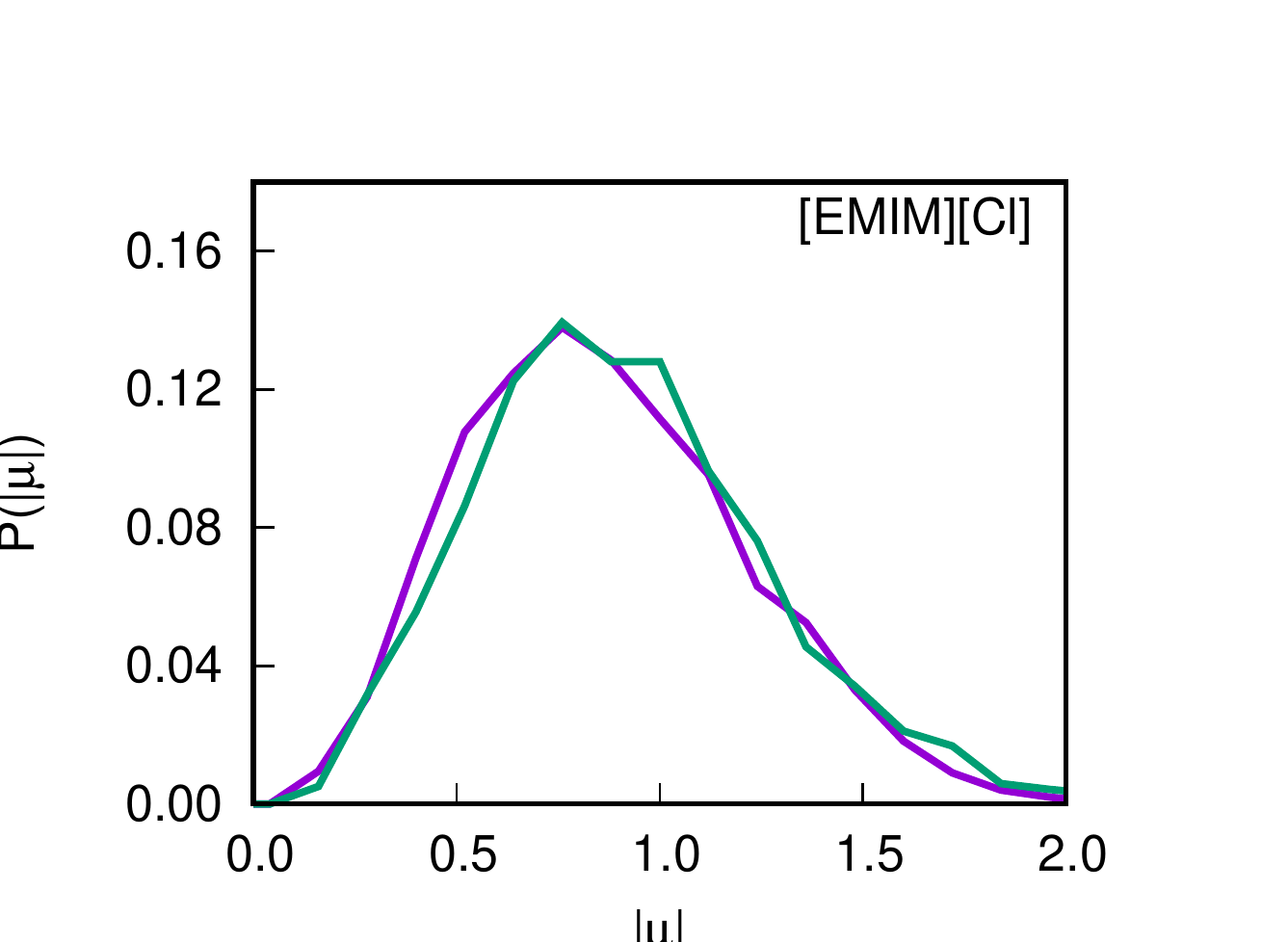}
                \includegraphics[clip=true,trim=2cm 2cm 6cm
                16cm,width=8cm]{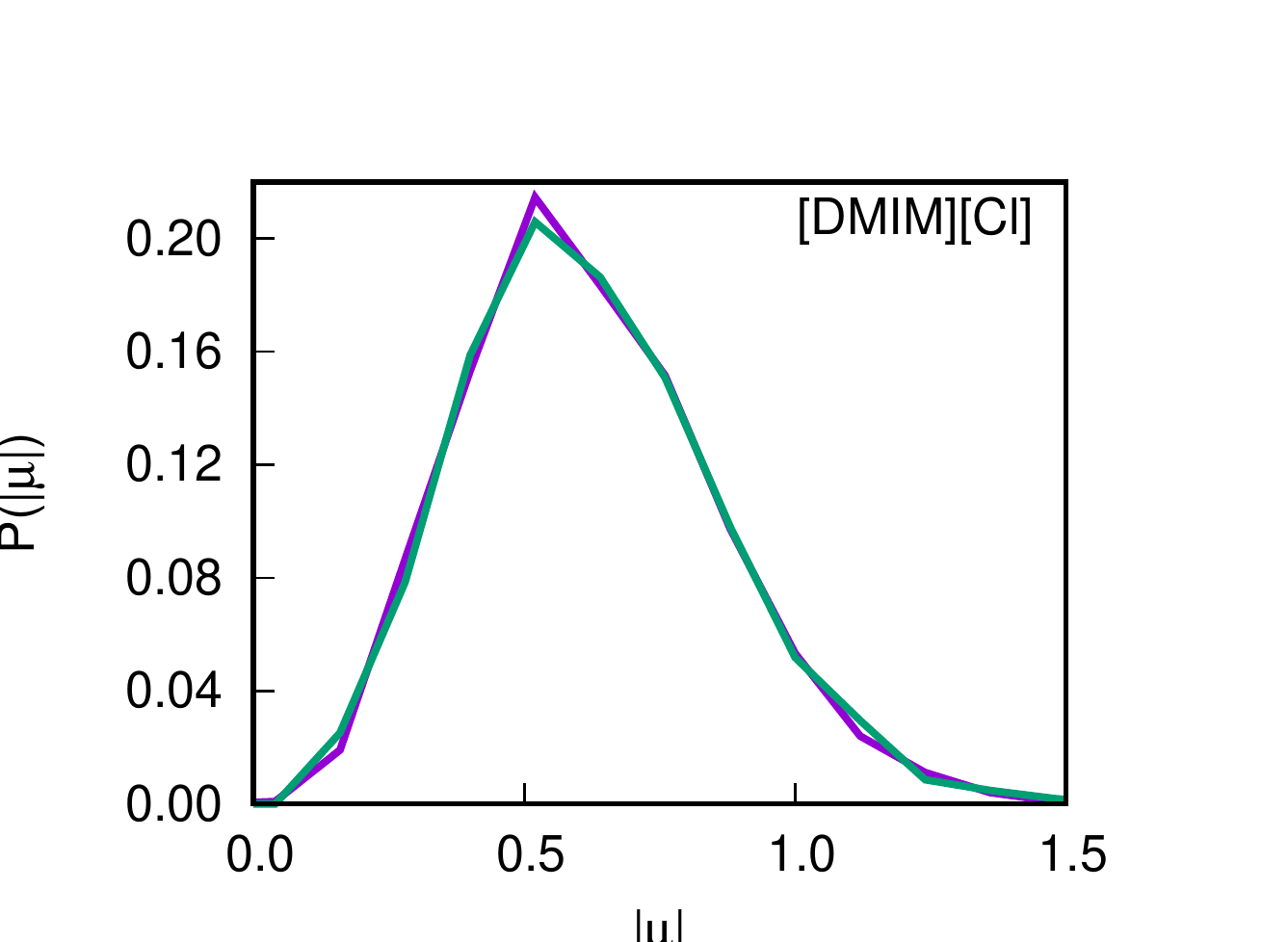}
                \caption{Probability distribution of the total dipole moment of
                the nanodroplet. The dipole moment, $\mu$  is calculated as
              $\sum_iq_i(dr_i)$. Where, $i$ - is the number of atoms of the
              molecular ions, $dr_i$ is the distance between the atoms and the
              center of the atomistic region or the box.}
                \label{fig:dm.il}
                \end{figure}

\begin{figure}[!htbp]
	        \centering
                \includegraphics[clip=true,trim=0cm 0cm 0cm
                0cm,width=12cm]{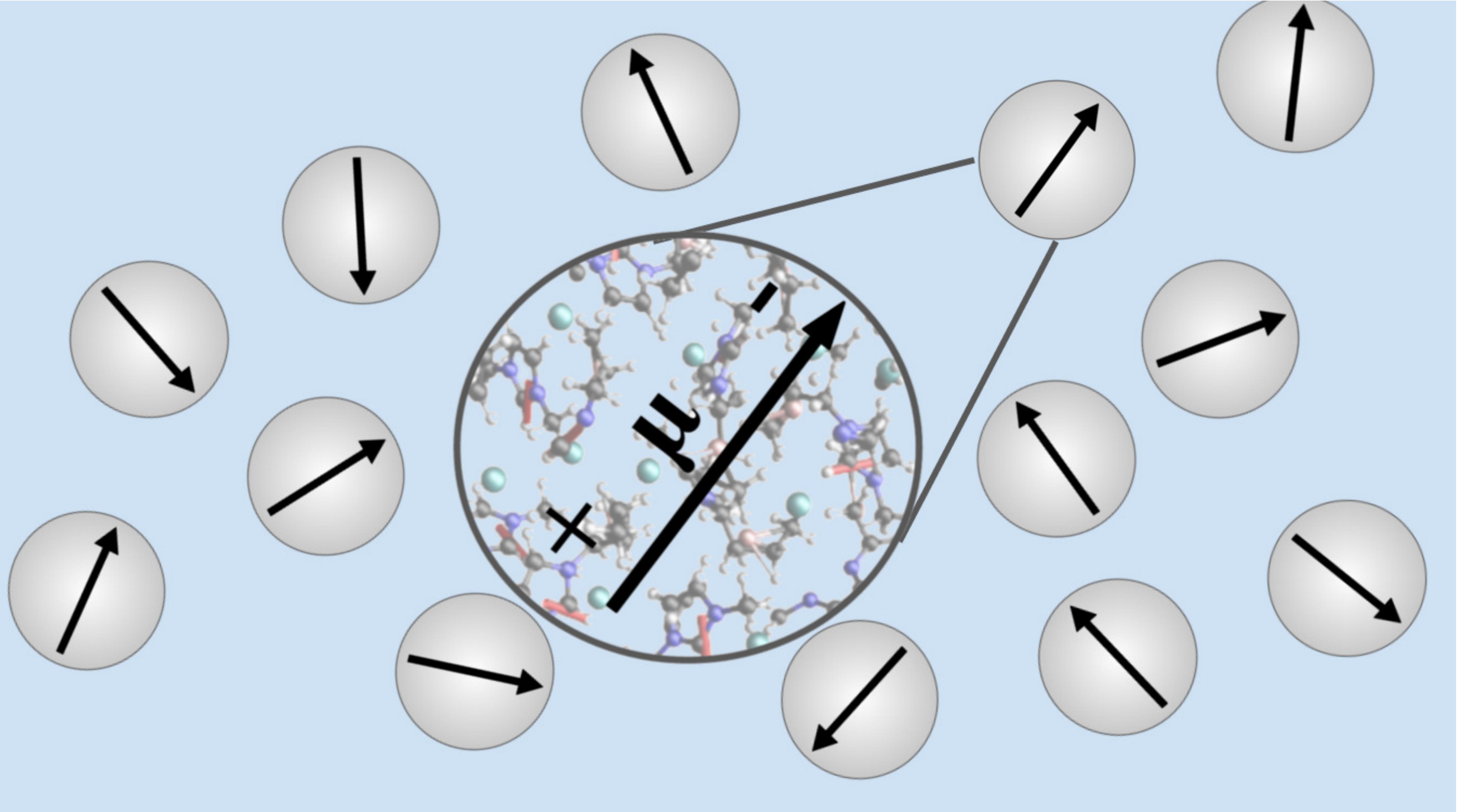}\\
                \caption{Cartoon representation of the nanodroplets
                embedded in a generic fluid}
                \label{fig:dipfig}
                \end{figure}

We can now
proceed to further explore the physical characteristics of the nanodroplet
defined by GC-AdResS. Being ILs strongly characterized as a mixture of positive and negative ions, a particularly
interesting quantity to look at is the (ensemble) average dipole moment of the
droplet. Although ILs display spatial heterogeneity at certain scale (see e.g.
\cite{wjyv07,turton2009}) in general one expects that on larger spatial scales there is a certain homogeneity, that is the nanodroplets are expected to be neutral (in average). However, within the single nanodroplet there may exist spatial heterogeneity, e.g. regions with higher concentration of anions (cations) thus the nanodroplets are characterized by an electric dipole moment. Indeed we have found that a picture of this kind can be drawn from our results.
Figs.\ref{fig:dm.il}, show the comparison between the AdResS results and the
same quantity calculated in the corresponding sphere in a full atomistic
simulation. The agreement is highly satisfactory and is within a minimum accuracy of $5\%$ (see also table \ref{tab:avg.dm}).
\begin{table}
\caption{Total dipole moment of the nanodroplet for each IL considered. It is reported also the systematic error obtained from AdResS and full atomistic simulations. The unit of $\mu$ is in Debye.}
\label{tab:avg.dm}       
\begin{tabular}{ccc}
\hline
\hline
System & AdResS & full atomistic\\
\hline
$[DMIM][Cl]$ & 0.56$\pm$0.005 & 0.55$\pm$0.005 \\
$[EMIM][Cl]$ & 0.84$\pm$0.007 & 0.81$\pm$0.005 \\
$[BMIM][Cl]$ & 0.83$\pm$0.005 & 0.86$\pm$0.005\\
$[BMIM][BF_4]$ & 0.98$\pm$0.008 & 1.02$\pm$0.009 \\
\hline
\hline
\end{tabular}
\end{table}
\begin{table}
\caption{Average number of ion pairs computed in the subregion of the AdResS and the same quantity computed in the equivalent subdomain of the full atomistic simulations}
\label{tab:avg.pn}       
\begin{tabular}{ccc}
\hline
\hline
System & full atomistic & AdResS\\
\hline
$[DMIM][Cl]$ & 582$\pm$2 & 594$\pm$2 \\
$[EMIM][Cl]$ & 565$\pm$2 & 554$\pm$2 \\
$[BMIM][Cl]$ & 137$\pm$1 & 130$\pm$1\\
$[BMIM][BF_4]$ & 336$\pm$2 & 343$\pm$1 \\
\hline
\hline
\end{tabular}
\end{table}
Interestingly the existence of anion-rich/cation-rich domains at smaller scale,
is consistent with results and conclusion of previous work
\cite{wv05,wv06,ur04,cp06} and once again it strengthen the idea of ILs as substances characterized by scales with different spatial
heterogeneity at mesoscopic level \cite{wjyv07,turton2009}. 
In perspective, one may think of a model for this class of ILs
as a collections of dipolar nanodroplets embedded in a generic (mean-field)
fluid, as sketched in Fig.\ref{fig:dipfig}, however, while it may represent an
interesting research plan, at this stage it goes beyond the scope of the current work.
Another interesting point is that the picture of the liquid as a collection of
dipolar nanodroplets may strengthen the complementarity in the interpretation of
relevant experimental results. In fact,
Israelachvili et {\it al.} \cite{gvb13} have shown that ionic liquids behave as weak diluted electrolytes where the electrostatic energy
between ions decay exponentially as a function of distance between them, that is, the Debye
screening or decay length monotonically decreases even in the limit of high
ion concentration because of the effective
screening of charges at shorter separations. In other words, long-range
forces are expected to not play a major role in concentrated electrolytes such
as ionic liquids (when we use the Debye screening model). At the same time, the
group of Perkin \cite{lpsp17,slp16} have shown that the conventional of strong
screening in ILs does not hold in the limit of high charge concentration of ions. Using simple scaling theory they show that there is an anomalous
dependence of the decay length on the ion concentration of an order of
magnitude larger than the ion diameter in ILs. Our results allow for an
interpretation that enforces the complementarity between these two views. In
fact, while the local structure of the droplet does not explicitly depend on the
electrostatic long range interactions (our coarse-grained model does not have
charges), however each nanodroplet is characterized by a dipole, that is the
larger spatial scale structure of the liquid (i.e. the structure of the liquid
as a collection of nanodroplets) instead will be influenced by the dipole-dipole
interaction of the nanodroplets (consistently with other proposals, see Refs.
\cite{susancomm,gvb13,gebbie15,slp16,rk16}). Moreover, it is important to note that the long screening length measured in Ref.\cite{slp16} is in accordance with inflated screening lengths in highly concentrated simple electrolytes such as 3M NaCl in water, thus it demonstrates that the details of molecular/ionic structures are probably not important for determining the long screening length. This conclusion can be drawn also from our study where beyond the local scale of the nanodroplet the specific chemical structure of the molecules does not play anymore a role, in fact we can use generic (and uncharged) spherical coarse-grained models. Finally, another important aspect related to the dipole of the
nanodroplet is that the unambiguous locality proven by AdResS is also
qualitatively supported by electronic structure-based related research.
In fact the vibrational spectra calculated from quantum simulations of
small samples are in very good agreement with experiments. One can
qualitatively conclude that indeed there may exist subregions of the
liquid which do not need the explicit environment for the calculation
of certain microscopic properties\cite{barbaravib}.
\section{Conclusion}
We have employed molecular dynamics within the GC-AdResS scheme and used a
statistical mechanics/probabilistic criterion to define and to quantify the size
of nanodomain in a class of imidazolium-based ILs. The main conclusion concerns the
characterization of the ILs as liquid formed by mesoscopic nanodroplets, a picture consistent
and complementary to the results obtained in other studies. Thus, in conclusion,
our aim is to enforce the idea, based on previous excellent work of other
groups, that molecular simulation can indeed be employed to defined nanodomains
in ILs. As a practical consequence, such studies can certainly help with a detailed {\it in silico} rational design of ions
for macroscopic properties on the liquid on demand.
\section{Methods}
Molecular Dynamics has been used for the simulation of the systems considered. In particular, within such an approach, we have used the Adaptive resolution simulation method (AdResS) \cite{adress2005,annurev2008} in its Grand Canonical (GC) set up \cite{wss12,whss13,azhws15,lds16,physrep2017}. The standard Molecular Dynamics method is well established and does not need further explanation, instead the AdResS method is not as popular and thus we report here its basic features that complement the description provided before in the text. A detailed description can be found in the references cited above.\\
In the Adaptive Resolution Simulation technique, space is divided in three regions, CG, HY and AT, characterized by coarse-grained, hybrid and atomistic resolution respectively and molecules adapt their structural resolution according to the region in which they are instantaneously located.
The algorithm is based on a  space dependent interpolation for the force between two molecules, $\alpha,\beta$ (see Fig.\ref{fig:cartoon.il}):
\begin{equation}
F_{\alpha \beta} = w(X_{\alpha})w(X_{\alpha})F_{\alpha\beta}^{AT} + [1 - w(X_{\alpha})w(X_{\alpha})]F_{\alpha\beta}^{CG}
\end{equation}  
$F_{\alpha\beta}^{AT}$ is the atomistic force while $F_{\alpha\beta}^{CG}$ is the coarse-grained force.  
$w(x)$ is a smooth function that goes from 0 to 1 in a transition region $\Delta$ (with $x$ center of mass of the molecule), a thermostat assures a target thermodynamic equilibrium. An additional force, ${\bf F}_{th}(x)$ (thermodynamic force), derived from first principles of statistical mechanics, acts on the center of mass of the molecule in $\Delta$. Such a force gives thermodynamic consistency to the method, i.e. assures that the effective chemical potential of the whole system corresponds to that of the (target) atomistic resolution. With this set up, the approach becomes numerically robust and conceptual rigorous for the simulation of system with open boundaries, thus in case the coarse-grained region is large enough, the method describes a Grand Canonical system for the atomistic region (GC-AdResS).
In practice, ${\bf F}_{th}(x)$ is calculated during a simulation (equilibration run), as the gradient of the number density of the molecules, in an iterative form \cite{prl2012}: $F_{k+1}^{th}(x)=F_{k}^{th}(x) - \frac{M_{\alpha}}{[\rho_{ref}]^2\kappa}\nabla\rho_{k}(x)$,
$M_{\alpha}$ is the mass of the molecule, $\kappa$ a tunable constant, $\rho_{k}(x)$ is the molecular density in $\Delta$, at the $k$-th iteration and $\rho_{ref}$ is the target density of reference.
The criterion of convergence depends on the accuracy required for the simulation, in general, $|\rho_{final}-\rho_{ref}|$ must be always below $10\%$ in $\Delta$. The GC-AdResS method has been implemented in the Molecular Dynamics package GROMACS \cite{hkvl08}, below are reported the specific details of the simulations of this work.\\
For each of the ILs, we have set up two simulations: 1. a full atomistic simulation
which we use it as a reference to compare the statistical properties calculated
in the sub-atomistic region of AdResS with the equivalent subregion of the full
atomistic simulations. 2. GC-AdResS simulation with spherical atomistic
resolution. All simulations were optimized using NPT ensemble. The temperature
and time step corresponds to 400K and 2fs respectively. The particle mesh Ewald
(PME) is used to determine the coulombic interactions. For the first 5ns we use Berendsen
barostat\cite{bpv84} which is then switched to Parrinello-Rahman\cite{pr81}
barostat for the remaining length of the simulation. 
The equilibration is carried over until the fluctuating box length reach an
order of 0.0001nm. This configuration is used as the initial configuration for
the production runs in NVT ensemble.All the statistical quantities were computed during the production run. The
production run length was 10ns for [DMIM][Cl] and 25-40ns for [BMIM][BF$_4$].
The force field
parameters, the step-by-step procedure to obtain the neutral coarse grained potentials
and the  thermodynamic forces for the coarse grained beads for
the four different IL systems are discussed elsewhere
\cite{jkkd18,jk18,ks17}. 
We use the tabulated CG potentials and thermodynamic force data to setup the GC-AdResS simulations. The GC-AdResS simulation with spherical resolution of the atomistic
region for two of the IL systems ([DMIM][Cl] and
[BMIM][BF$_4$]) are discussed here and the reminder systems are discussed
elsewhere\cite{jkkd18,jk18}.
The radius of the atomistic region in the GC-AdResS simulation is fixed at 3.0nm
for both [DMIM][Cl] and [BMIM][BF$_4$] (see table \ref{tab:locality.charge}),
which turned to be the minimal size of the atomistic region that reproduces all the possible atom-atom radial distribution functions. The hybrid region that borders the sub atomic region has
a width of 2nm for each system. The coarse grained region spreads out throughout
the reminder of the cubic box of length 15.69333\AA\ (20000 ion pairs [DMIM][Cl]) and
14.17269(8451 ion pairs [BMIM][BF$_4$])
respectively. An optimized $\Gamma = 5
ps^{-1}$ parameter is used with Langevin thermostat. Reaction field method is used to
calculate the long ranged electrostatic interaction between
particles\cite{fjk12}. The quality of the thermodynamic force is such that the
density in the atomistic subregion of the GC-AdResS is comparable to the bulk
atomistic system with a maximum fluctuations $<5\%$ for both the systems.
\section{Acknowledgment}
This research has been funded by Deutsche Forschungsgemeinschaft (DFG)
through the grant CRC 1114: ``Scaling Cascades in Complex Systems'',
project C01. We would like to thank  Prof.Barbara Kirchner and Prof.Susan Perkin for relevant suggestions that shaped the work.

\providecommand{\latin}[1]{#1}
\providecommand*\mcitethebibliography{\thebibliography}
\csname @ifundefined\endcsname{endmcitethebibliography}
  {\let\endmcitethebibliography\endthebibliography}{}


\begin{mcitethebibliography}{50}
\providecommand*\natexlab[1]{#1}
\providecommand*\mciteSetBstSublistMode[1]{}
\providecommand*\mciteSetBstMaxWidthForm[2]{}
\providecommand*\mciteBstWouldAddEndPuncttrue
  {\def\EndOfBibitem{\unskip.}}
\providecommand*\mciteBstWouldAddEndPunctfalse
  {\let\EndOfBibitem\relax}
\providecommand*\mciteSetBstMidEndSepPunct[3]{}
\providecommand*\mciteSetBstSublistLabelBeginEnd[3]{}
\providecommand*\EndOfBibitem{}
\mciteSetBstSublistMode{f}
\mciteSetBstMaxWidthForm{subitem}{(\alph{mcitesubitemcount})}
\mciteSetBstSublistLabelBeginEnd
  {\mcitemaxwidthsubitemform\space}
  {\relax}
  {\relax}

\bibitem[Liu \latin{et~al.}(2008)Liu, Luo, Wu, Liu, Zhang, and Chen]{lllzc08}
Liu,~N.; Luo,~F.; Wu,~H.; Liu,~Y.; Zhang,~C.; Chen,~J. {One-Step
  Ionic-Liquid-Assisted Electrochemical Synthesis of
  Ionic-Liquid-Functionalized Graphene Sheets Directly from Graphite}.
  \emph{Adv. Funct. Mater.} \textbf{2008}, \emph{18}, 1518--1525\relax
\mciteBstWouldAddEndPuncttrue
\mciteSetBstMidEndSepPunct{\mcitedefaultmidpunct}
{\mcitedefaultendpunct}{\mcitedefaultseppunct}\relax
\EndOfBibitem
\bibitem[Zhang \latin{et~al.}(2010)Zhang, Ning, Zhang, Qiao, Zhang, He, and
  Liu]{znzqzhl10}
Zhang,~B.; Ning,~W.; Zhang,~J.; Qiao,~X.; Zhang,~J.; He,~J.; Liu,~C.-Y. {Stable
  Dispersions of Reduced Graphene Oxide in Ionic Liquids}. \emph{J. Mater.
  Chem.} \textbf{2010}, \emph{20}, 5401--5403\relax
\mciteBstWouldAddEndPuncttrue
\mciteSetBstMidEndSepPunct{\mcitedefaultmidpunct}
{\mcitedefaultendpunct}{\mcitedefaultseppunct}\relax
\EndOfBibitem
\bibitem[Fukushima \latin{et~al.}(2003)Fukushima, Kosaka, Ishimura, Yamamoto,
  Takigawa, Ishii, and Aida]{fki03}
Fukushima,~T.; Kosaka,~A.; Ishimura,~Y.; Yamamoto,~T.; Takigawa,~T.; Ishii,~N.;
  Aida,~T. {Molecular Ordering of Organic Molten Salts Triggered by
  Single-Walled Carbon Nanotubes}. \emph{Science} \textbf{2003}, \emph{300},
  2072--2074\relax
\mciteBstWouldAddEndPuncttrue
\mciteSetBstMidEndSepPunct{\mcitedefaultmidpunct}
{\mcitedefaultendpunct}{\mcitedefaultseppunct}\relax
\EndOfBibitem
\bibitem[Li \latin{et~al.}(2014)Li, Chen, Wu, He, and Jin]{lcwhj14}
Li,~H.; Chen,~L.; Wu,~H.; He,~H.; Jin,~Y. {Ionic Liquid-Functionalized
  Fluorescent Carbon Nanodots and Their Applications in Electrocatalysis,
  Biosensing, and Cell Imaging}. \emph{Langmuir} \textbf{2014}, \emph{30},
  15016--15021\relax
\mciteBstWouldAddEndPuncttrue
\mciteSetBstMidEndSepPunct{\mcitedefaultmidpunct}
{\mcitedefaultendpunct}{\mcitedefaultseppunct}\relax
\EndOfBibitem
\bibitem[Sham and Notley(2013)Sham, and Notley]{sn13}
Sham,~A. Y.~W.; Notley,~S.~M. {A Review of Fundamental Properties and
  Applications of Polymer-Graphene Hybrid Materials}. \emph{Soft Matter}
  \textbf{2013}, \emph{9}, 6645--6653\relax
\mciteBstWouldAddEndPuncttrue
\mciteSetBstMidEndSepPunct{\mcitedefaultmidpunct}
{\mcitedefaultendpunct}{\mcitedefaultseppunct}\relax
\EndOfBibitem
\bibitem[Gindri \latin{et~al.}(2014)Gindri, Frizzo, Bender, Tier, Martins,
  Villetti, Machado, Rodriguez, and Rodrigues]{gfb14}
Gindri,~I.~M.; Frizzo,~C.~P.; Bender,~C.~R.; Tier,~A.~Z.; Martins,~M. A.~P.;
  Villetti,~M.~A.; Machado,~G.; Rodriguez,~L.~C.; Rodrigues,~D.~C. {Preparation
  of TiO$_2$ Nanoparticles Coated with Ionic Liquids: A Supramolecular
  Approach}. \emph{ACS Appl. Mater. Interfaces} \textbf{2014}, \emph{6},
  11536--11543\relax
\mciteBstWouldAddEndPuncttrue
\mciteSetBstMidEndSepPunct{\mcitedefaultmidpunct}
{\mcitedefaultendpunct}{\mcitedefaultseppunct}\relax
\EndOfBibitem
\bibitem[Gutel \latin{et~al.}(2009)Gutel, Santini, Philippot, Padua, Pelzer,
  Chaudret, Chauvin, and Basset]{gsp09}
Gutel,~T.; Santini,~C.~C.; Philippot,~K.; Padua,~A.; Pelzer,~K.; Chaudret,~B.;
  Chauvin,~Y.; Basset,~J.-M. {Organized 3D-Alkyl Imidazolium Ionic Liquids
  Could Be Used to Control the Size of ${\it in situ}$ Generated Ruthenium
  Nanoparticles?} \emph{J. Mater. Chem.} \textbf{2009}, \emph{19},
  3624--3631\relax
\mciteBstWouldAddEndPuncttrue
\mciteSetBstMidEndSepPunct{\mcitedefaultmidpunct}
{\mcitedefaultendpunct}{\mcitedefaultseppunct}\relax
\EndOfBibitem
\bibitem[Urahata and Ribeiro(2004)Urahata, and Ribeiro]{ur04}
Urahata,~S.~M.; Ribeiro,~M. C.~C. {Structure of Ionic Liquids of 1-
  Alkyl-3-Methylimidazolium Cations: A Systematic Computer Simulation Study}.
  \emph{J. Chem. Phys.} \textbf{2004}, \emph{120}, 1855--1863\relax
\mciteBstWouldAddEndPuncttrue
\mciteSetBstMidEndSepPunct{\mcitedefaultmidpunct}
{\mcitedefaultendpunct}{\mcitedefaultseppunct}\relax
\EndOfBibitem
\bibitem[Wang and Voth(2005)Wang, and Voth]{wv05}
Wang,~Y.; Voth,~G.~A. {Unique Spatial Heterogeneity in Ionic Liquids}. \emph{J.
  Am. Chem. Soc.} \textbf{2005}, \emph{127}, 12192--12193\relax
\mciteBstWouldAddEndPuncttrue
\mciteSetBstMidEndSepPunct{\mcitedefaultmidpunct}
{\mcitedefaultendpunct}{\mcitedefaultseppunct}\relax
\EndOfBibitem
\bibitem[Canongia~Lopes and Padua(2006)Canongia~Lopes, and Padua]{cp06}
Canongia~Lopes,~J. N.~A.; Padua,~A. A.~H. {Nanostructural Organization in Ionic
  Liquids}. \emph{J. Phys. Chem. B} \textbf{2006}, \emph{110}, 3330--3335\relax
\mciteBstWouldAddEndPuncttrue
\mciteSetBstMidEndSepPunct{\mcitedefaultmidpunct}
{\mcitedefaultendpunct}{\mcitedefaultseppunct}\relax
\EndOfBibitem
\bibitem[Wang and Voth(2006)Wang, and Voth]{wv06}
Wang,~Y.; Voth,~G.~A. {Tail Aggregation and Domain Diffusion in Ionic Liquids}.
  \emph{J. Phys. Chem. B} \textbf{2006}, \emph{110}, 18601--18608\relax
\mciteBstWouldAddEndPuncttrue
\mciteSetBstMidEndSepPunct{\mcitedefaultmidpunct}
{\mcitedefaultendpunct}{\mcitedefaultseppunct}\relax
\EndOfBibitem
\bibitem[Triolo \latin{et~al.}(2007)Triolo, Russina, Bleif, and
  Di~Cola]{trbd07}
Triolo,~A.; Russina,~O.; Bleif,~H.-J.; Di~Cola,~E. {Nanoscale Segregation in
  Room Temperature Ionic Liquids}. \emph{J. Phys. Chem. B} \textbf{2007},
  \emph{111}, 4641--4644\relax
\mciteBstWouldAddEndPuncttrue
\mciteSetBstMidEndSepPunct{\mcitedefaultmidpunct}
{\mcitedefaultendpunct}{\mcitedefaultseppunct}\relax
\EndOfBibitem
\bibitem[Busetti \latin{et~al.}(2010)Busetti, Crawford, Earle, Gilea, Gilmore,
  Gorman, Laverty, Lowry, and McLaughlin]{bce10}
Busetti,~A.; Crawford,~D.~E.; Earle,~M.~J.; Gilea,~M.~A.; Gilmore,~B.~F.;
  Gorman,~S.~P.; Laverty,~G.; Lowry,~A.~F.; McLaughlin,~S. K.~R.,~M.;
  {Antimicrobial and Antibiofilm Activities of 1-Alkylquinolinium Bromide Ionic
  Liquids.} \emph{Green Chem.} \textbf{2010}, \emph{12}, 420--425\relax
\mciteBstWouldAddEndPuncttrue
\mciteSetBstMidEndSepPunct{\mcitedefaultmidpunct}
{\mcitedefaultendpunct}{\mcitedefaultseppunct}\relax
\EndOfBibitem
\bibitem[Kumar and Malhotra(2009)Kumar, and Malhotra]{km09}
Kumar,~V.; Malhotra,~S.~V. {Study on the Potential Anti-Cancer Activity of
  Phosphonium and Ammonium-Based Ionic Liquids.} \emph{Bioorg. Med. Chem.
  Lett.} \textbf{2009}, \emph{19}, 4643--4646\relax
\mciteBstWouldAddEndPuncttrue
\mciteSetBstMidEndSepPunct{\mcitedefaultmidpunct}
{\mcitedefaultendpunct}{\mcitedefaultseppunct}\relax
\EndOfBibitem
\bibitem[Angell and Zhao(2013)Angell, and Zhao]{az13}
Angell,~C.~A.; Zhao,~Z. {Fluctuations, Clusters, and Phase Transitions in
  Liquids, Solutions, and Glasses: from Metastable Water to Phase Change Memory
  Materials}. \emph{Faraday Discuss} \textbf{2013}, \emph{167}, 625--641\relax
\mciteBstWouldAddEndPuncttrue
\mciteSetBstMidEndSepPunct{\mcitedefaultmidpunct}
{\mcitedefaultendpunct}{\mcitedefaultseppunct}\relax
\EndOfBibitem
\bibitem[Hayes \latin{et~al.}(2015)Hayes, Warr, and Atkin]{hwa15}
Hayes,~R.; Warr,~G.~G.; Atkin,~R. {Structure and Nanostructure in Ionic
  Liquids}. \emph{Chem. Rev.} \textbf{2015}, \emph{115}, 6357--6426\relax
\mciteBstWouldAddEndPuncttrue
\mciteSetBstMidEndSepPunct{\mcitedefaultmidpunct}
{\mcitedefaultendpunct}{\mcitedefaultseppunct}\relax
\EndOfBibitem
\bibitem[Wendler \latin{et~al.}(2011)Wendler, Zahn, Dommert, Berger, Holm,
  Kirchner, and {Delle Site}]{wzdbhkd11}
Wendler,~K.; Zahn,~S.; Dommert,~F.; Berger,~R.; Holm,~C.; Kirchner,~B.; {Delle
  Site},~L. {Locality and Fluctuations: Trends in Imidazolium-Based Ionic
  Liquids and Beyond}. \emph{J. Chem. Theory Comput.} \textbf{2011}, \emph{7},
  3040--3044\relax
\mciteBstWouldAddEndPuncttrue
\mciteSetBstMidEndSepPunct{\mcitedefaultmidpunct}
{\mcitedefaultendpunct}{\mcitedefaultseppunct}\relax
\EndOfBibitem
\bibitem[Gozzo \latin{et~al.}(2004)Gozzo, Santos, Augusti, Consorti, Dupont,
  and Eberlin]{gsacde04}
Gozzo,~F.~C.; Santos,~L.~S.; Augusti,~R.; Consorti,~C.~S.; Dupont,~J.;
  Eberlin,~M.~N. {Gaseous Supramolecules of Imidazolium Ionic Liquids: "Magic"
  Numbers and Intrinsic Strengths of Hydrogen Bonds}. \emph{Chem. Eur. J.}
  \textbf{2004}, \emph{10}, 6187--6193\relax
\mciteBstWouldAddEndPuncttrue
\mciteSetBstMidEndSepPunct{\mcitedefaultmidpunct}
{\mcitedefaultendpunct}{\mcitedefaultseppunct}\relax
\EndOfBibitem
\bibitem[Poblete \latin{et~al.}(2010)Poblete, Praprotnik, Kremer, and {Delle
  Site}]{jcp2010-simon}
Poblete,~S.; Praprotnik,~M.; Kremer,~K.; {Delle Site},~L. {Coupling Different
  Levels of Resolution in Molecular Simulations}. \emph{J. Chem. Phys.}
  \textbf{2010}, \emph{132}, 114101\relax
\mciteBstWouldAddEndPuncttrue
\mciteSetBstMidEndSepPunct{\mcitedefaultmidpunct}
{\mcitedefaultendpunct}{\mcitedefaultseppunct}\relax
\EndOfBibitem
\bibitem[Agarwal and {Delle Site}(2015)Agarwal, and {Delle Site}]{jcp-2015-pi}
Agarwal,~A.; {Delle Site},~L. {Path Integral Molecular Dynamics Within the
  Grand Canonical-Like Adaptive Resolution Technique: Simulation of Liquid
  Water}. \emph{J. Chem. Phys.} \textbf{2015}, \emph{143}, 094102\relax
\mciteBstWouldAddEndPuncttrue
\mciteSetBstMidEndSepPunct{\mcitedefaultmidpunct}
{\mcitedefaultendpunct}{\mcitedefaultseppunct}\relax
\EndOfBibitem
\bibitem[Lambeth \latin{et~al.}(2010)Lambeth, Junghans, Kremer, Clementi, and
  {Delle Site}]{jcp-2010-full}
Lambeth,~B.; Junghans,~C.; Kremer,~K.; Clementi,~C.; {Delle Site},~L. {On the
  Locality of Hydrogen Bond Networks at Hydrophobic Interfaces}. \emph{J. Chem.
  Phys.} \textbf{2010}, \emph{133}, 221101\relax
\mciteBstWouldAddEndPuncttrue
\mciteSetBstMidEndSepPunct{\mcitedefaultmidpunct}
{\mcitedefaultendpunct}{\mcitedefaultseppunct}\relax
\EndOfBibitem
\bibitem[Praprotnik \latin{et~al.}(2008)Praprotnik, {Delle Site}, and
  Kremer]{annurev2008}
Praprotnik,~M.; {Delle Site},~L.; Kremer,~K. {Multiscale Simulation of Soft
  Matter: From Scale Bridging to Adaptive Resolution}. \emph{Annu. Rev. Phys.
  Chem.} \textbf{2008}, \emph{59}, 545--571\relax
\mciteBstWouldAddEndPuncttrue
\mciteSetBstMidEndSepPunct{\mcitedefaultmidpunct}
{\mcitedefaultendpunct}{\mcitedefaultseppunct}\relax
\EndOfBibitem
\bibitem[Wang \latin{et~al.}(2013)Wang, Hartmann, Sch\"{u}tte, and {Delle
  Site}]{whss13}
Wang,~H.; Hartmann,~C.; Sch\"{u}tte,~C.; {Delle Site},~L. {Grand-Canonical-Like
  Molecular-Dynamics Simulations by Using an Adaptive-Resolution Technique}.
  \emph{Phys. Rev. X} \textbf{2013}, \emph{3}, 011018\relax
\mciteBstWouldAddEndPuncttrue
\mciteSetBstMidEndSepPunct{\mcitedefaultmidpunct}
{\mcitedefaultendpunct}{\mcitedefaultseppunct}\relax
\EndOfBibitem
\bibitem[Agarwal \latin{et~al.}(2015)Agarwal, Zhu, Hartmann, Wang, and {Delle
  Site}]{azhws15}
Agarwal,~A.; Zhu,~J.; Hartmann,~C.; Wang,~H.; {Delle Site},~L. {Molecular
  Dynamics in a Grand Ensemble: Bergmann-Lebowitz Model and Adaptive Resolution
  Simulation}. \emph{New J. Phys.} \textbf{2015}, \emph{17}, 083042\relax
\mciteBstWouldAddEndPuncttrue
\mciteSetBstMidEndSepPunct{\mcitedefaultmidpunct}
{\mcitedefaultendpunct}{\mcitedefaultseppunct}\relax
\EndOfBibitem
\bibitem[Fritsch \latin{et~al.}(2012)Fritsch, Poblete, Junghans, Ciccotti,
  {Delle Site}, and Kremer]{prl2012}
Fritsch,~S.; Poblete,~S.; Junghans,~C.; Ciccotti,~G.; {Delle Site},~L.;
  Kremer,~K. {Adaptive Resolution Molecular Dynamics Simulation Through
  Coupling to an Internal Particle Reservoir}. \emph{Phys. Rev. Lett.}
  \textbf{2012}, \emph{108}, 170602\relax
\mciteBstWouldAddEndPuncttrue
\mciteSetBstMidEndSepPunct{\mcitedefaultmidpunct}
{\mcitedefaultendpunct}{\mcitedefaultseppunct}\relax
\EndOfBibitem
\bibitem[Delle~Site(2016)]{lds16}
Delle~Site,~L. {Formulation of Liouville's Theorem for Grand Ensemble Molecular
  Simulations}. \emph{Phys. Rev. E} \textbf{2016}, \emph{93}, 022130\relax
\mciteBstWouldAddEndPuncttrue
\mciteSetBstMidEndSepPunct{\mcitedefaultmidpunct}
{\mcitedefaultendpunct}{\mcitedefaultseppunct}\relax
\EndOfBibitem
\bibitem[Dommert \latin{et~al.}(2012)Dommert, Wendler, Bergerm, {Delle Site},
  and Holm]{dwbsh12}
Dommert,~F.; Wendler,~K.; Bergerm,~R.; {Delle Site},~L.; Holm,~C. {Force Fields
  for Studying the Structure and Dynamics of Ionic Liquids: A Critical Review
  of Recent Developments}. \emph{ChemPhysChem} \textbf{2012}, \emph{13},
  1625\relax
\mciteBstWouldAddEndPuncttrue
\mciteSetBstMidEndSepPunct{\mcitedefaultmidpunct}
{\mcitedefaultendpunct}{\mcitedefaultseppunct}\relax
\EndOfBibitem
\bibitem[Lehmann \latin{et~al.}(2010)Lehmann, Roatsch, Schoppke, and
  Kirchner]{lrsk10}
Lehmann,~S. B.~C.; Roatsch,~M.; Schoppke,~M.; Kirchner,~B. {On the Physical
  Origin of the Cation-Anion Intermediate Bond in Ionic Liquids Part I. Placing
  a (Weak) Hydrogen Bond Between Two Charges}. \emph{Phys. Chem. Chem. Phys.}
  \textbf{2010}, \emph{12}, 7473--7486\relax
\mciteBstWouldAddEndPuncttrue
\mciteSetBstMidEndSepPunct{\mcitedefaultmidpunct}
{\mcitedefaultendpunct}{\mcitedefaultseppunct}\relax
\EndOfBibitem
\bibitem[Agarwal and {Delle Site}(2017)Agarwal, and {Delle
  Site}]{pccp-2017-full}
Agarwal,~C.,~A.and~Clementi; {Delle Site},~L. {Path Integral-GC-AdResS
  Simulation of a Large Hydrophobic Solute in Water: A Tool to Investigate the
  Interplay Between Local Microscopic Structures and Quantum Delocalization of
  Atoms in Space }. \emph{Phys. Chem. Chem. Phys.} \textbf{2017}, \emph{19},
  13030--13037\relax
\mciteBstWouldAddEndPuncttrue
\mciteSetBstMidEndSepPunct{\mcitedefaultmidpunct}
{\mcitedefaultendpunct}{\mcitedefaultseppunct}\relax
\EndOfBibitem
\bibitem[Shadrack~Jabes \latin{et~al.}(2018)Shadrack~Jabes, Klein, and
  Delle~Site]{jkd18}
Shadrack~Jabes,~B.; Klein,~R.; Delle~Site,~L. {Structural Locality and Early
  Stage of Aggregation of Micelles in Water: An Adaptive Resolution Molecular
  Dynamics Study}. \emph{Adv. Theory Simul.} \textbf{2018}, \emph{1},
  1800025\relax
\mciteBstWouldAddEndPuncttrue
\mciteSetBstMidEndSepPunct{\mcitedefaultmidpunct}
{\mcitedefaultendpunct}{\mcitedefaultseppunct}\relax
\EndOfBibitem
\bibitem[Shadrack~Jabes \latin{et~al.}(2018)Shadrack~Jabes, Krekeler, Klein,
  and Delle~Site]{jkkd18}
Shadrack~Jabes,~B.; Krekeler,~C.; Klein,~R.; Delle~Site,~L. {Probing Spatial
  Locality in Ionic Liquids with the Grand Canonical Adaptive Resolution
  Molecular Dynamics Technique}. \emph{J. Chem. Phys.} \textbf{2018},
  \emph{148}, 193804\relax
\mciteBstWouldAddEndPuncttrue
\mciteSetBstMidEndSepPunct{\mcitedefaultmidpunct}
{\mcitedefaultendpunct}{\mcitedefaultseppunct}\relax
\EndOfBibitem
\bibitem[Shadrack~Jabes and Krekeler(2018)Shadrack~Jabes, and Krekeler]{jk18}
Shadrack~Jabes,~B.; Krekeler,~C. {Ionic Liquids Treated within the Grand
  Canonical Adaptive Resolution Molecular Dynamics Technique}.
  \emph{Computation} \textbf{2018}, \emph{6}, 23\relax
\mciteBstWouldAddEndPuncttrue
\mciteSetBstMidEndSepPunct{\mcitedefaultmidpunct}
{\mcitedefaultendpunct}{\mcitedefaultseppunct}\relax
\EndOfBibitem
\bibitem[Krekeler and {Delle Site}(2017)Krekeler, and {Delle Site}]{ks17}
Krekeler,~C.; {Delle Site},~L. {Towards Open Boundary Molecular Dynamics
  Simulation of Ionic Liquids}. \emph{Phys. Chem. Chem. Phys.} \textbf{2017},
  \emph{19}, 4701--4709\relax
\mciteBstWouldAddEndPuncttrue
\mciteSetBstMidEndSepPunct{\mcitedefaultmidpunct}
{\mcitedefaultendpunct}{\mcitedefaultseppunct}\relax
\EndOfBibitem
\bibitem[Wang \latin{et~al.}(2007)Wang, Jiang, Yan, and Voth]{wjyv07}
Wang,~Y.; Jiang,~W.; Yan,~T.; Voth,~G.~A. {Understanding Ionic Liquids Through
  Atomistic and Coarse-Grained Molecular Dynamics Simulations}. \emph{Acc.
  Chem. Res.} \textbf{2007}, \emph{40}, 1193--1199\relax
\mciteBstWouldAddEndPuncttrue
\mciteSetBstMidEndSepPunct{\mcitedefaultmidpunct}
{\mcitedefaultendpunct}{\mcitedefaultseppunct}\relax
\EndOfBibitem
\bibitem[Turton \latin{et~al.}(2009)Turton, Hunger, Stoppa, Hefter, A.~Thoman,
  Buchner, and Wynne]{turton2009}
Turton,~D.~A.; Hunger,~J.; Stoppa,~A.; Hefter,~G.; A.~Thoman,~M.~W.;
  Buchner,~R.; Wynne,~K. {Dynamics of Imidazolium Ionic Liquids from a Combined
  Dielectric Relaxation and Optical Kerr Effect Study: Evidence for Mesoscopic
  Aggregation}. \emph{J. Am. Chem. Soc.} \textbf{2009}, \emph{131}, 11140\relax
\mciteBstWouldAddEndPuncttrue
\mciteSetBstMidEndSepPunct{\mcitedefaultmidpunct}
{\mcitedefaultendpunct}{\mcitedefaultseppunct}\relax
\EndOfBibitem
\bibitem[Gebbie \latin{et~al.}(2013)Gebbie, Valtiner, Banquy, Fox, Henderson,
  and Israelachvili]{gvb13}
Gebbie,~M.~A.; Valtiner,~M.; Banquy,~X.; Fox,~E.~T.; Henderson,~W.~A.;
  Israelachvili,~J.~N. {Ionic Liquids Behave as Dilute Electrolyte Solutions}.
  \emph{P. Natl. Acad. Sci. U. S. A} \textbf{2013}, \emph{110},
  9674--9679\relax
\mciteBstWouldAddEndPuncttrue
\mciteSetBstMidEndSepPunct{\mcitedefaultmidpunct}
{\mcitedefaultendpunct}{\mcitedefaultseppunct}\relax
\EndOfBibitem
\bibitem[Lee \latin{et~al.}(2017)Lee, Perez-Martinez, Smith, and
  Perkin]{lpsp17}
Lee,~A.~A.; Perez-Martinez,~C.~S.; Smith,~A.~M.; Perkin,~S. {Scaling Analysis
  of the Screening Length in Concentrated Electrolytes}. \emph{Phys. Rev.
  Lett.} \textbf{2017}, \emph{119}, 026002\relax
\mciteBstWouldAddEndPuncttrue
\mciteSetBstMidEndSepPunct{\mcitedefaultmidpunct}
{\mcitedefaultendpunct}{\mcitedefaultseppunct}\relax
\EndOfBibitem
\bibitem[Smith \latin{et~al.}(2016)Smith, Lee, and Perkin]{slp16}
Smith,~A.~M.; Lee,~A.~A.; Perkin,~S. {The Electrostatic Screening Length in
  Concentrated Electrolytes Increases with Concentration}. \emph{J. Phys. Chem.
  Lett} \textbf{2016}, \emph{7}, 2157--2163\relax
\mciteBstWouldAddEndPuncttrue
\mciteSetBstMidEndSepPunct{\mcitedefaultmidpunct}
{\mcitedefaultendpunct}{\mcitedefaultseppunct}\relax
\EndOfBibitem
\bibitem[Gebbie \latin{et~al.}(2017)Gebbie, Smith, Dobbs, Lee, Warr, Banquy,
  Valtiner, Rutland, Israelachvili, Perkin, , and Atkin]{susancomm}
Gebbie,~M.~A.; Smith,~A.~M.; Dobbs,~H.~A.; Lee,~A.~A.; Warr,~G.~G.; Banquy,~X.;
  Valtiner,~M.; Rutland,~M.~W.; Israelachvili,~J.~N.; Perkin,~S. \latin{et~al.}
   {Long Range Electrostatic Forces in Ionic Liquids}. \emph{Chem. Commun.}
  \textbf{2017}, \emph{53}, 1214\relax
\mciteBstWouldAddEndPuncttrue
\mciteSetBstMidEndSepPunct{\mcitedefaultmidpunct}
{\mcitedefaultendpunct}{\mcitedefaultseppunct}\relax
\EndOfBibitem
\bibitem[Gebbie \latin{et~al.}(2015)Gebbie, Dobbs, Valtiner, and
  Israelachvili]{gebbie15}
Gebbie,~M.~A.; Dobbs,~H.~A.; Valtiner,~M.; Israelachvili,~J.~N. {Long-Range
  Electrostatic Screening in Ionic Liquids}. \emph{Proc. Natl. Acad. Sci. U. S.
  A.} \textbf{2015}, \emph{112}, 7432--7437\relax
\mciteBstWouldAddEndPuncttrue
\mciteSetBstMidEndSepPunct{\mcitedefaultmidpunct}
{\mcitedefaultendpunct}{\mcitedefaultseppunct}\relax
\EndOfBibitem
\bibitem[Kjellander(2016)]{rk16}
Kjellander,~R. {Decay Behavior of Screened Electrostatic Surface Forces in
  Ionic Liquids: The Vital Role of Non-Local Electrostatics}. \emph{Phys. Chem.
  Chem. Phys.} \textbf{2016}, \emph{18}, 18985--19000\relax
\mciteBstWouldAddEndPuncttrue
\mciteSetBstMidEndSepPunct{\mcitedefaultmidpunct}
{\mcitedefaultendpunct}{\mcitedefaultseppunct}\relax
\EndOfBibitem
\bibitem[Thomas \latin{et~al.}(2014)Thomas, Brehm, Holloczki, Kelemen,
  Nyulaszi, Pasinszki, and Kirchner]{barbaravib}
Thomas,~M.; Brehm,~M.; Holloczki,~O.; Kelemen,~Z.; Nyulaszi,~L.; Pasinszki,~T.;
  Kirchner,~B. {Simulating the Vibrational Spectra of Ionic Liquid Systems:
  1-Ethyl-3-Methylimidazolium Acetate and Its Mixtures}. \emph{J. Chem. Phys.}
  \textbf{2014}, \emph{141}, 024510\relax
\mciteBstWouldAddEndPuncttrue
\mciteSetBstMidEndSepPunct{\mcitedefaultmidpunct}
{\mcitedefaultendpunct}{\mcitedefaultseppunct}\relax
\EndOfBibitem
\bibitem[Praprotnik \latin{et~al.}(2005)Praprotnik, {Delle Site}, and
  Kremer]{adress2005}
Praprotnik,~M.; {Delle Site},~L.; Kremer,~K. {Adaptive Resolution
  Molecular-Dynamics Simulation: Changing the Degrees of Freedom on the Fly}.
  \emph{J. Chem. Phys.} \textbf{2005}, \emph{123}, 224106\relax
\mciteBstWouldAddEndPuncttrue
\mciteSetBstMidEndSepPunct{\mcitedefaultmidpunct}
{\mcitedefaultendpunct}{\mcitedefaultseppunct}\relax
\EndOfBibitem
\bibitem[Wang \latin{et~al.}(2012)Wang, Sch\"{u}tte, and {Delle Site}]{wss12}
Wang,~H.; Sch\"{u}tte,~C.; {Delle Site},~L. {Adaptive Resolution Simulation
  (AdResS): A Smooth Thermodynamic and Structural Transition from Atomistic to
  Coarse Grained Resolution and Vice Versa in a Grand Canonical Fashion}.
  \emph{J. Chem. Theory Comput.} \textbf{2012}, \emph{8}, 2878--2887\relax
\mciteBstWouldAddEndPuncttrue
\mciteSetBstMidEndSepPunct{\mcitedefaultmidpunct}
{\mcitedefaultendpunct}{\mcitedefaultseppunct}\relax
\EndOfBibitem
\bibitem[{Delle Site} and Praprotnik(2017){Delle Site}, and
  Praprotnik]{physrep2017}
{Delle Site},~L.; Praprotnik,~M. {Molecular Systems with Open Boundaries:
  Theory and Simulation}. \emph{Phys. Rep.} \textbf{2017}, \emph{693},
  1--56\relax
\mciteBstWouldAddEndPuncttrue
\mciteSetBstMidEndSepPunct{\mcitedefaultmidpunct}
{\mcitedefaultendpunct}{\mcitedefaultseppunct}\relax
\EndOfBibitem
\bibitem[Hess \latin{et~al.}(2008)Hess, Kutzner, van~der Spoel, and
  Lindahl]{hkvl08}
Hess,~B.; Kutzner,~C.; van~der Spoel,~D.; Lindahl,~E. {GROMACS 4: Algorithms
  for Highly Efficient, Load-Balanced, and Scalable Molecular Simulation}.
  \emph{J. Chem. Theory Comput.} \textbf{2008}, \emph{4}, 435\relax
\mciteBstWouldAddEndPuncttrue
\mciteSetBstMidEndSepPunct{\mcitedefaultmidpunct}
{\mcitedefaultendpunct}{\mcitedefaultseppunct}\relax
\EndOfBibitem
\bibitem[Berendsen \latin{et~al.}(1984)Berendsen, Postma, van Gunsteren,
  DiNola, and Haak]{bpv84}
Berendsen,~H. J.~C.; Postma,~J. P.~M.; van Gunsteren,~W.~F.; DiNola,~A.;
  Haak,~J.~R. {Molecular Dynamics with Coupling to an External Bath}. \emph{J.
  Chem. Phys.} \textbf{1984}, \emph{81}, 3684--3690\relax
\mciteBstWouldAddEndPuncttrue
\mciteSetBstMidEndSepPunct{\mcitedefaultmidpunct}
{\mcitedefaultendpunct}{\mcitedefaultseppunct}\relax
\EndOfBibitem
\bibitem[Parrinello and Rahman(1981)Parrinello, and Rahman]{pr81}
Parrinello,~M.; Rahman,~A. {Polymorphic Transitions in Single Crystals: A New
  Molecular Dynamics Method}. \emph{J. Appl. Phys.} \textbf{1981}, \emph{52},
  7182\relax
\mciteBstWouldAddEndPuncttrue
\mciteSetBstMidEndSepPunct{\mcitedefaultmidpunct}
{\mcitedefaultendpunct}{\mcitedefaultseppunct}\relax
\EndOfBibitem
\bibitem[Fritsch \latin{et~al.}(2012)Fritsch, Junghans, and Kremer]{fjk12}
Fritsch,~S.; Junghans,~C.; Kremer,~K. {Structure Formation of Toluene Around
  C60: Implementation of the Adaptive Resolution Scheme (AdResS) into GROMACS}.
  \emph{J. Chem. Theory Comput.} \textbf{2012}, \emph{8}, 398--403\relax
\mciteBstWouldAddEndPuncttrue
\mciteSetBstMidEndSepPunct{\mcitedefaultmidpunct}
{\mcitedefaultendpunct}{\mcitedefaultseppunct}\relax
\EndOfBibitem
\end{mcitethebibliography}
\end{document}